\documentclass[onecolumn,useAMS,usegraphicx,usenatbib,aas_macros]{mn2e}
\usepackage{array}
\usepackage{amsfonts}
\usepackage{amssymb}
\usepackage{mathrsfs}
\usepackage{amsmath}
\usepackage{subfigure}
\defcitealias{Rudiger_Kitchatinov:1996}{RK96}
\defcitealias{Charbonneau_MacGregor:1993}{CM93}

\title[Angular momentum transport in radiative interiors]{A semi-analytic approach to angular momentum transport in stellar radiative interiors}
\author[F. Spada et al.]
{  F.~Spada$^{1,2}$,
 A.C.~Lanzafame$^{1,2}$ and
 A.F.~Lanza$^2$  \\
$^1$ Sezione Astrofisica, Dipartimento di Fisica e Astronomia, Universit\`a degli Studi di Catania, Via S. Sofia, 78, 95123, Catania, Italy
\\
$^2$ INAF - Osservatorio Astrofisico di Catania, Via S. Sofia, 78, 95123, Catania, Italy }

\begin{document}

\date{}

\pagerange{\pageref{firstpage}--\pageref{lastpage}} \pubyear{}

\maketitle

\label{firstpage}

\begin{abstract}
We address the problem of angular momentum transport in stellar radiative interiors with a novel semi-analytic spectral technique, using an eigenfunction series expansion, that can be used to derive benchmark solutions in hydromagnetic regimes with very high Reynolds number ($10^{7} - 10^{8}$).
The error arising from the truncation of the series is evaluated analytically.
The main simplifying assumptions are the neglect of meridional circulation and of non-axisymmetric magnetic fields. The advantages of our approach are shown by applying it to a spin-down model for a $1\,M_\odot$ main-sequence star.
The evolution of the coupling between core and envelope is investigated for different values of the viscosity  and different geometries and values of the poloidal field.
We confirm that a viscosity enhancement by $\sim 10^{4}$ with respect to the molecular value is required to attain a rigid rotation in the core of the Sun within its present age.
We suggest that a quadrupolar poloidal field may explain the short coupling  time-scale needed to model the observed rotational evolution of fast rotators on the ZAMS, while a dipolar geometry is indicated in the case of slow rotators.
Our novel semi-analytic spectral method provides  a conceptually simple and rigorous treatment of a classic MHD problem and allows us to explore the influence of various parameters on the rotational history of radiative interiors.
\end{abstract}

\begin{keywords}
MHD -- methods: analytical -- methods: numerical -- stars: rotation -- stars: magnetic fields -- stars: late-type
\end{keywords}

\section{Introduction}
\label{intro}

The  rotational evolution of a solar type star during the main sequence lifetime is deeply affected by the level of internal coupling that is established between the radiative core and the convective envelope at various ages, which competes with the angular momentum loss produced by a magnetised stellar wind. In spite of the  observational constraints gathered so far, the identification of the physical process(es)  ensuring a uniform rotation of the core, i.e., a  substantial rotational coupling by the age of the Sun, is still lacking.

Observational information  to constrain the rotational evolution of low mass stars currently comes from two major sources: helioseismology and  rotation period surveys in open clusters. The former provides evidence of a latitudinal dependence of solar rotation in the convection zone (CZ), the existence of a thin shear layer (tachocline) at the top of the radiative core, and an almost uniform rotation at greater depths \citep[at least down to $0.2 - 0.3$ $R_\odot$, see][]{Thompson_ea:2003}. On the other hand, rotation period measurements of open cluster members effectively constrain the rotational evolution because of our fairly reliable estimate of their age. A satisfactory theoretical explanation of these data is still missing owing to the limited knowledge of the different processes involved in the evolution of stellar angular momentum.

During the pre-main sequence (PMS) contraction phase, stars are spun up, but contemporarily suffer a remarkable angular momentum draining by the interaction with their circumstellar discs \citep{Koenigl:1991}. Very young (viz., a few Myr old) open clusters already show significant spread and mass-dependent structures in their rotation period distributions \citep{Lamm_ea:2004}. On the main sequence (MS), angular momentum loss by a magnetised wind in solar-type stars \citep{Weber_Davis:1967,Kawaler:1988} was early recognised by \citet{Schatzman:1962} as the main cause of their rotational evolution, contrasting with the more massive stars' behaviour \citep[see][]{Kraft:1967,Skumanich:1972}.

Surface period measurements are sensitive to the internal rotational profile.
Phenomenological models of MS angular momentum evolution \citep{Allain:1998, Bouvier:2008b} have  been  successful in reproducing the observations if a certain amount of differential rotation between the radiative core and the convective envelope is allowed. In those models the exchange of angular momentum between the core and the envelope is governed by a single parameter, i.e., a coupling time-scale $\tau_{\rm c}$ which fixes the rate at which angular momentum is transferred between the two regions to establish rigid rotation. The observed distribution of the rotation periods in open clusters can be reproduced by assuming that $\tau_{\rm c}$ is of the order of $\sim 10$ Myr for the stars that begun their evolution on the ZAMS as fast rotators (i.e., with an initial  rotation period of a few days), while it is remarkably longer ($\tau_{\rm c} \sim 100$ Myr) for slow rotators (i.e., with initial periods of several days). An understanding of the processes eventually ensuring rotational coupling between core and envelope and within the core itself is still lacking.

The issue is also closely related to a number of currently open questions in stellar evolution theory. For instance, \citet{Bouvier:2008b} tentatively explained the observed correlation between Li over-depletion and the presence of hot Jupiters around MS stars on the basis of these rotational evolution models because an initial slow rotation is a characteristic of stars with exoplanets, owing to their prolonged interaction with a proto-planetary disc during their PMS phase. Such objects are characterized by a long coupling time ($\tau_{\rm c} \sim 100$ Myr) which leads to the development of a sizeable differential rotation at the core-envelope interface. A sizeable shear produces an enhancement of the turbulent mixing which in turn makes Li destruction more efficient.
To address the internal coupling processes from a theoretical point of view, \citet{Spruit:1999, Spruit:2002} discussed the stability of toroidal magnetic fields in radiative stellar interiors. He  suggested that, due to the vertically stabilizing effect of the subadiabatic stratification of the core, motions associated with hydro-magnetic instabilities are constrained to the locally horizontal direction and their mean effect can be described through an enhancement of the momentum and magnetic field diffusivities up to turbulent values several orders of magnitude greater than their molecular counterparts.

\citet{Charbonneau_MacGregor:1993}, hereafter \citetalias{Charbonneau_MacGregor:1993}, solved the equations for angular momentum transport and toroidal magnetic field evolution in a solar-type star throughout the MS lifespan, assuming axial symmetry and a pre-existent (fossil) poloidal magnetic field. They used a finite elements technique, finding that a sizeable toroidal field develops from the winding-up of the poloidal field by the differential rotation. When it becomes strong enough, the Maxwell stresses begin to react to any further amplification of the field and the system enters into a new regime characterized by torsional Alfv\'en oscillations. This phase displays a remarkable energy dissipation  as soon as the oscillations along neighbour magnetic field lines get in opposition of phase (phase-mixing), owing to their slightly different periods, as a consequence of the density stratification and poloidal field gradient inside the core \citep{Spruit:1999}.  \citetalias{Charbonneau_MacGregor:1993} find that the poloidal field geometry is an  important feature to determine the time-scale for angular momentum redistribution inside the core, because phase mixing (much more efficiently than diffusion) enhances angular momentum redistribution and toroidal field reconnection on surfaces of constant poloidal field, thus producing a quasi-stationary regime with an angular velocity almost constant on them, according to Ferraro isorotation theorem \citep{Mestel_ea:1988}. Another key finding is the impossibility to achieve a state of uniform rotation of the core within the solar age with molecular viscosity alone. Later work by \citet{Rudiger_Kitchatinov:1996} (\citetalias{Rudiger_Kitchatinov:1996}), based on a finite difference numerical approach, confirmed these results and remarkably the "viscosity deficit" problem.

We solve the same coupled, non-homogeneous partial differential equations (PDEs) of \citetalias{Charbonneau_MacGregor:1993} and \citetalias{Rudiger_Kitchatinov:1996} for angular momentum transport and toroidal field evolution, with the technique of  eigenfunction expansion \citep{Morse_Feshbach:1953, Haberman:2004}. We find an exact analytic solution of the problem that can be expressed as a series expansion. Formal substitution of the series into the equations yields an infinite system of linear, first order ordinary differential equations (ODEs) that can be truncated according to the required degree of accuracy. To this purpose, we complement our implementation with an analytic formula for  the truncation error.
In principle, our spectral method has no limitation in terms of Reynolds number and can be used as a benchmark to test the accuracy of other numerical methods to solve the same problem.
This new rigorous treatment of a classic MHD problem is the novelty reported in the present work. We shall compare the results of our approach with those of previous works by \citetalias{Charbonneau_MacGregor:1993} and \citetalias{Rudiger_Kitchatinov:1996}, briefly discussing its advantages for the study of the angular momentum transport in a radiative stellar core.
Our treatment is particularly relevant for an accurate study of the phase mixing process for which previous numerical techniques were not capable of  a rigorous treatment of the contribution of subgrid lengthscales.

\section{Model}
\label{sec:math}

We model the MS evolution of specific angular momentum in the core of a 1\,$M_{\odot}$ star solving the angular momentum transport equation and the toroidal component of the induction equation.

\subsection{Basic assumptions}
\label{basic_ass}

We consider an inertial reference frame with the origin $O$ at the barycentre of the star and the polar axis $\bf z$ along the rotation axis. Spherical polar coordinates are adopted, $r$ being the distance from $O$, $\vartheta$ the colatitude measured from the North pole, and $\varphi$ the azimuthal angle.
 
The present calculation is based on a number of simplifying assumptions. We restrict our domain to the radiative core, assuming that the angular momentum transport time-scale in the CZ is much shorter than in the radiative interior, as turbulent viscosity there is at least $10 -12$ orders of magnitude greater than molecular. The system is assumed to be strictly axisymmetric.

The \citet{Spitzer:1962} expressions for molecular viscosity and magnetic diffusivity are:
\begin{equation}
\begin{array}{ll}
\nu_\mathrm{mol}  =& 1.2 \cdot 10^{-16} \ {T^{5/2}}{\rho}^{-1} \mbox{~cm$^{2}$ s$^{-1}$~}, \\
\eta_\mathrm{mol} =& 10^{13} \ T^{-3/2} \mbox{~cm$^{2}$ s$^{-1}$~}.
\end{array}
\label{eq:diff}
\end{equation}

As pointed out by \citetalias{Rudiger_Kitchatinov:1996}, molecular viscosity is far too low to reconcile the differential rotation regime that is likely established in young, solar-type stars with the uniform rotation  of the core of the present Sun as deduced by helioseismology. To overcome this "viscosity deficit" problem, following \citetalias{Rudiger_Kitchatinov:1996}, we introduce an artificial viscosity enhancement parameter $R_\mathrm{eff}$ such that:
\begin{equation}
\nu_\mathrm{eff} = R_\mathrm{eff} \nu_\mathrm{mol}.
\label{eq:visc_enhance}
\end{equation}
This suffices to the purpose of presenting our method and comparing the results with previous works. \citet{Spruit:1999} and \citet{Denissenkov_Pinsonneault:2007} present a possible identification and physical explanation for such an effect.

Application of Eqs.\,(\ref{eq:diff}) requires the knowledge of the stellar structure (viz. the depth dependence of density and temperature). Here we use the so-called model S of the Sun introduced by \citet{CDea:1996} for the whole computation.
The  evolution of the stellar structure on the MS has a minor impact on our computations and can be safely ignored.

We neglect  the Eddington-Sweet circulation because of the extremely long time-scale that is of the order of $10^{12}$ yr. Note that a differential rotation which is not uniform along cylindrical surfaces around the $\bf z$ axis drives a meridional circulation because of the non-potential character of the associated centrifugal force. We shall neglect such a circulation because the subadiabatic stratification of the core effectively opposes motions in the radial direction, strongly reducing its velocity.  

Our assumptions rule out the possibility of poloidal field regeneration by dynamo action and 3D  magnetic instabilities. According to the analysis of magnetohydrodynamic (MHD) instabilities in radiative regions by \citet{Spruit:1999}, however, the fastest-growing instability should be the $m=1$ mode of the Tayler instability, which may be included as an additional Maxwell stress term in our equations.

Our assumption of an axisymmetric poloidal magnetic field is justified by the fact that  in a differentially rotating core any initially non-axisymmetric field component  is smoothed out by winding up and  diffusion  on a time-scale $\tau_{\rm w}$ much shorter than  the diffusion time-scale of the axisymmetric component.  \citet[][ Sect. 3.1]{Spruit:1999} estimates  $\tau_{\rm w}$ of the order of $10^{2}$ yr for the Sun on the ZAMS. Note that if the initial field is so strong to oppose the winding up by the initial differential rotation, it can possibly find a non-axisymmetric equilibrium \citep{BraithwaiteSpruit04}. However, in this case the core would be rotating rigidly from the outset being completely coupled by the
strong field and the angular momentum would be transported on the Alfven time scale which is of the order of $10^{2}-10^{4}$ yr, i.e., much shorter than the wind braking time scale. We do not treat this case because such a strong coupling inside a young solar-like star is not in agreement with the phenomenological models of rotational evolution requiring coupling time scales of $10^{7}-10^{8}$ yr to account for the observations, as discussed in Sect.~\ref{intro}.

\subsection{Governing equations}

With the hypotheses discussed above, the total fluid velocity $\bf u$ and magnetic field $\bf B$ can be written as:
\begin{eqnarray*}
{\bf u} &=& r\sin \vartheta \Omega(r, \vartheta,t) {\bf e}_\varphi, \\
{\bf B} &=& {\bf B}_p + {\bf B}_t =  \frac{1}{r\sin \vartheta} \left[ \frac{1}{r} \frac{\partial \Psi(r,\vartheta,t)}{\partial \vartheta} {\bf e}_r -  \frac{\partial \Psi(r,\vartheta,t)}{\partial r} {\bf e}_\vartheta \right] + B_\varphi(r,\vartheta,t) {\bf e}_\varphi.
\end{eqnarray*}
The axisymmetric magnetic field is expressed in terms of two scalar functions, the flux function $\Psi$ and the toroidal component $B_\varphi$. The poloidal field lines lie on surfaces of constant $\Psi$ (magnetic surfaces) as it is immediately apparent from ${\bf B}_p \cdot \nabla \Psi  = 0$.  

The evolution of $\Omega$ is governed by the angular momentum conservation law \citep{Rudiger:1989}:
\begin{equation}
\label{eq:omega}
\rho r^2 \sin^2 \vartheta \frac{\partial \Omega}{\partial t} = \nabla \cdot \left( \rho r^2 \sin^2 \vartheta \nu \nabla \Omega \right) + \frac{1}{4\pi} {\bf B}_p \cdot \nabla (r \sin \vartheta B_\varphi),  
\end{equation}
and that of $B_\varphi$ and $\Psi$ by the induction equation, which breaks into two scalar equations \citep{Radler:1980}:
\begin{align}
\label{eq:field}
\frac{\partial B_\varphi}{\partial t} &= \left[ \nabla^2 - \frac{1}{r^2 \sin^2 \vartheta} \right] B_\varphi + \frac{\nabla \eta \cdot \nabla (r \sin \vartheta B_\varphi)}{r \sin \vartheta} + r \sin \vartheta {\bf B}_p \cdot \nabla \Omega,  
\\
\label{eq:psi}
\frac{\partial \Psi}{\partial t} &= \eta \frac{\partial^2 \Psi }{\partial r^2} + \eta \frac{\sin \vartheta}{r^2} \frac{\partial }{\partial \vartheta} \left( \frac{1}{\sin \vartheta} \frac{\partial \Psi}{\partial \vartheta} \right).
\end{align}

The evolution of $\Psi$, as given by Eq.\,\ref{eq:psi}, is decoupled from Eqs. \ref{eq:omega} and \ref{eq:field}, but it requires knowledge of initial conditions which are generally not available \citepalias[cf. ][]{Rudiger_Kitchatinov:1996}.

The presence of a magnetic field within the core is expected as a relic of dynamo action during PMS. This problem has been addressed by \citet{Kitchatinov_ea:2001} for stars like the Sun. Their numerical calculations go from an age of a few Myr to about $30$ Myr and account for the dynamical retreat of the CZ during the PMS evolution. They find that non-axisymmetric modes are the most readily excited in young PMS models, but they are gradually replaced by  axisymmetric ones as the star approaches the ZAMS.
Moreover, any  early non-axisymmetric field is rapidly diffused away, as already noted at the end of Sect.~\ref{basic_ass}.  We shall therefore consider a stationary, axisymmetric configuration for the poloidal seed field, assigned through a flux function $\Psi$, and specialize our considerations to the case in which $\Psi$ can be factorized as $\Psi (r, \theta) = \tilde{\varphi}(r) \chi(\theta)$. Specifically, we investigate the case of a dipole
confined within the spherical shell $r_{1} \leq r \leq r_{2}$ (see Sect.~\ref{BCandI}) with:
\begin{equation}
\Psi = {B_{0}}{R_*^2} \left(r-r_{1} \right)^{2} \left(r-r_{2} \right)^{2}  \sin^2 \vartheta,
\label{eq:dipstream}
\end{equation}
where $B_{0}$ is the scale of the field intensity and $R_*$ is the radius of the star. The next multipole, studied for comparison purpose, is the quadrupole, defined by:
\begin{equation}
\Psi =  {B_{0}}{R_*^2} \left(r - r_{1}\right)^{2} \left(r - r_{2} \right)^{2}  \sin \vartheta \cos \vartheta.
\label{eq:quadstream}
\end{equation}

\subsection{Boundary and initial conditions}
\label{BCandI}

Thanks to the assumption of axisymmetry, our computational domain encompasses a meridional section of the radiative core: $(r,\vartheta) \in [r_1,r_2] \times [0,\pi ]$; $r_2$ represents the lower boundary of the CZ (for the Sun, $r_2=0.7$ $R_\odot$), while $r_1$ is a nonzero lower boundary, introduced to avoid  singularity at the origin.
The actual value of $r_1$ does not significantly affect the results, provided that it is chosen in such a way that the cylinder where $ r \sin \theta \leq r_{1}$  contains a negligible amount of angular momentum.
We are interested in the global response of the core to an external wind braking torque, which we take into account through a suitable specification of the boundary conditions  at $r_2$.

Specifically, on the radial boundaries we assign the angular momentum fluxes. At the inner
boundary $r_1$, we assume it to be negligible. At the outer boundary $r_2$, the angular
momentum flux is equal to that lost via the magnetised wind:
\begin{equation}
\label{eq:om_bc}
\left. \frac{\partial \Omega}{\partial r}\right|_{r_1} = 0 \quad ; \quad \left. \frac{\partial \Omega}{\partial r}\right|_{r_2} = \mathscr{W}(t).
\end{equation}
For simplicity, we neglect any dependence of the wind torque and $\partial \Omega/\partial r$ on the latitude, which is in any case a second-order effect, given the high turbulent viscosity of the CZ which makes the amplitude of the latitudinal shear on top of the boundary significantly smaller than that in the core.
For the specification of the function $\mathscr{W}(t)$, which includes a model of the wind braking process, see Appendix \ref{app:wind}.

For the toroidal magnetic field we use ``insulating" boundary conditions:
\begin{equation}
\label{eq:fl_bc}
B_\varphi|_{r_1}=0 \quad ; \quad B_\varphi|_{r_2}=0.
\end{equation}
They prevent the development of toroidal magnetic fields of unrealistically large intensities, as shown by, e.g., \citet{Garaud_Guervilly:2008} in numerical simulations of the solar tachocline. It is interesting to note that tachocline models predict the existence of a circulation inside that layer that confines the interior magnetic field preventing its outward diffusion \citep[e.g., ][]{SpiegelZahn:1992,GoughMcIntyre98,Gough07}. Nevertheless, even if the poloidal field diffused outward, the strong radial shear present in the tachocline would wind it up producing a strong azimuthal field.
In the mildly subadiabatic environment of the tachocline, it would become unstable and emerge on a time scale of $100-1000$ days  by, e.g., doubly diffusive instabilities \citep{SchmittRosner83,Caligarietal95,Silversetal09,Silversetal09b}. Therefore, such a toroidal flux will be rapidly removed from the upper boundary of our computational domain on a timescale so short as to justify the assumption that $B_{\varphi} = 0$ at the outer boundary.

Initial conditions for the problem at hand are the outcome of the PMS evolution, namely, the contraction of stellar radius from several to about one solar radii and the development of a convectively stable core. Although the angular velocity profile emerging from those processes is not known, we may assume that they establish an internal differential rotation with a gradient mainly in the radial direction. Therefore, we take as initial conditions:
\begin{eqnarray}
\label{eq:omega_ic}
\frac{\Omega_\mathrm{in}(r)}{\Omega_0} &=& 1 + \frac{\Delta \Omega}{\Omega_0} \left[ \frac{1}{2} + \frac{1}{\pi}{\rm atan} \left(\frac{ r - r_{\rm step}}{\Delta r_{\rm step}}\right) \right], \\
\label{eq:bphi_ic}
B_{\varphi,\mathrm{in}}&=& 0,
\end{eqnarray}
where $\Delta \Omega$ is the amplitude of a smooth step in the initial rotation profile $\Omega_\mathrm{in}(r)$, centered at $r_{\rm step}$ and with a width $\Delta r_{\rm step}$ (see Sect.\,\ref{sec:results}), $\Omega_0 \simeq 2.7 \cdot 10^{-6}$ s$^{-1}$ is the present solar angular velocity, used as a reference scale, and $B_{\varphi,\rm{in}}$ is the initial toroidal magnetic field.

These assumptions for the initial conditions should be regarded as working hypotheses, merely chosen for the sake of simplicity. In any case, they have little effect on the solution, as it is the driving wind torque at the outer boundary $r_2$ that dominates the subsequent angular momentum evolution after a brief transient phase.

\section{Solution method}
\label{sec:method}

The method used to solve the coupled PDEs \eqref{eq:omega} and \eqref{eq:field} is based on a novel, semi-analytic approach, relying on an eigenfunction expansion technique \citep[see, for example,][]{Haberman:2004}. Both Eqs. \eqref{eq:omega} and \eqref{eq:field} can be put in the form:
\begin{equation}
\label{eq:sample}
w(r,\vartheta) \frac{\partial}{\partial t}  u(r,\vartheta,t) - \mathcal{L}[u(r,\vartheta,t)] = Q(r,\vartheta,t),
\end{equation}
i.e. they consist of a linear homogeneous part (the l.h.s.), containing a differential operator $\cal L$ acting on the main dependent variable $u$, and of a r.h.s. source term (here called $Q$). $\cal L$ expresses the action of the spatial derivatives appearing in Eqs. \eqref{eq:omega} and \eqref{eq:field}.
Note that the source term in Eq. \eqref{eq:omega}  depends on the assigned function $\Psi$ and  the other dependent variable $B_\varphi$. Similarly, the source term in Eq. \eqref{eq:field} contains $\Omega$, thus producing a coupling between the two equations. Focusing on the homogeneous PDEs associated to Eqs. \eqref{eq:omega} and \eqref{eq:field}, the technique of separation of variables is applicable to each of them. First, we determine the eigenvalues $\kappa_n$ and eigenfunctions $\phi_n(r,\vartheta)$ of the linear differential operators $\cal L$ satisfying  homogeneous boundary conditions, i.e., we solve the Sturm-Liouville problems determined by:
\begin{equation}
\label{eq:sampeig}
{\cal L}[\phi_{n}] + \kappa_{n} w\ \phi_{n} = 0,
\end{equation}
and $\dfrac{\partial \phi_n}{\partial r} = 0$ for Eq. \eqref{eq:omega}, or $\phi_{n} = 0$ for Eq. \eqref{eq:field} on the radial boundaries $r_1$, $r_2$ and suitable regularity conditions on the $\bf z$ axis.
The function $w$ that appears as a factor of the time derivative in the l.h.s. of  Eq.~\eqref{eq:sample} enters the eigenvalue equation \eqref{eq:sampeig} as a weighting function (in our specific case, $w \equiv \rho r^2 \sin^2 \vartheta$ for Eq.~\eqref{eq:omega} and $w\equiv 1$ for Eq.~\eqref{eq:field}).
If the differential operator $\cal L$  is self-adjoint (as it is the case for Eqs.  \eqref{eq:omega} and \eqref{eq:field}, see Appendix \ref{app:selfadj} for details), the  solution $u(r,\vartheta,t)$ can be represented as a series expansion in terms of the eigenfunctions $\phi_{n}$, i.e.:
\begin{equation*}
u(r,\vartheta,t) = \sum_{n=0}^\infty a_{n}(t)\phi_{n}(r,\vartheta),
\end{equation*}
where the coefficients $a_{n}(t)$ are functions of the time.
In this way, the eigenfunctions provide the spatial dependence for the solution and warrant that it  satisfies the homogeneous boundary conditions.

The non-homogeneous source term $Q$, as well as the effects of non-homogeneous boundary conditions (as, in our case, the second of Eqs. \eqref{eq:om_bc}), are taken into account through similar series expansions in terms of the same eigenfunctions because they form a complete set, as it is known from the theory of the Sturm-Liouville problem. In such a way, a system of linear ODEs for the evolution of the expansion coefficients $a_{n}(t)$ is derived by a direct substitution of the series expansions into the original complete equations or, more formally, by applying the Green  identity, as it is explained in App. \ref{app:selfadj}. This yields:
\begin{equation*}
\frac{d a_{n}}{dt} + \kappa_{n}a_{n}(t) = b_{n}(t) + [\dots],
\end{equation*}
where the terms in the r.h.s. indicated by the dots are those coming from the expansion of the non-homogeneous contributions, i.e., $Q$ and the boundary conditions.

\subsection{Scaling of the equations}

Fixing a proper scaling for our variables is not a simple matter because of the remarkably different time-scales involved. Stellar rotation naturally sets the first time-scale:
\begin{equation*}
t_\Omega=2\pi\Omega_*^{-1},
\end{equation*}
where $\Omega_*$ is the mean surface angular velocity. Note that, for the whole MS, $\Omega_*$ is between $1 - 10$ $\Omega_\odot$, thus giving $t_\Omega \sim 10^{-2} - 10^{-1}$ yr.
This is also the characteristic time of toroidal field winding up and amplification by differential rotation.
The Alfv\'en velocity $u_{\rm A}$ introduces the Alfv\'en crossing time-scale:
\begin{eqnarray*}
u_{\rm A}=\frac{B_0}{\sqrt{4\pi\rho_0}}, \\
t_{\rm A}=R_*/u_{\rm A},
\end{eqnarray*}
where $\rho_0$ is the value of the density at the top of the radiative zone (i.e., $r=r_{2}$), chosen as a reference (numerically, $t_{\rm A} \sim 10^3$ yr). In addition, {\it global} time-scales for momentum and magnetic field diffusion arise in our problem:
\begin{eqnarray*}
t_\nu = R_*^2/\nu_0, \\
t_\eta = R_*^2/\eta_0,
\end{eqnarray*}
with values of $\sim 10^{12}$ and $\sim 10^{10}$ yr, respectively,
when the reference values of $\nu_0$ and $\eta_0$ are taken at the top of the radiative zone.

It should be stressed that $t_{\rm A}$, $t_\eta$ and $t_\nu$ are introduced here only for scaling purposes and are not representative of the time-scales of specific processes occurring in our system.

The best compromise between such different time-scales  is to choose $t_{\rm A}$ as the unit of time and construct nondimensional quantities as follows: $t^*=t/t_{\rm A}$, $r^*=r/R_*$, ${\bf B}_p^*={\bf B}_p/B_0$, $\Omega^*=\Omega t_{\rm A}$, $B_\varphi^*=B_\varphi/B_0$ 
\citepalias[cf.][]{Charbonneau_MacGregor:1993}. Dropping the asterisk superscripts on dimensionless quantities, the scaled equations become:
\begin{eqnarray}
\label{eq:om_sc}
\rho r^2 \sin^2 \vartheta \frac{\partial \Omega}{\partial t} - \frac{1}{\mathscr{R}_\nu} \nabla \cdot ( \rho r^2 \sin^2 \vartheta \nu \nabla \Omega) &=& {\bf B}_p \cdot \nabla (r \sin \vartheta B_\varphi), \\
\label{eq:fl_sc}
\frac{\partial B_\varphi}{\partial t} -\frac{1}{\mathscr{R}_\eta} \left\{ \left[ \nabla^2 - \frac{1}{r^2 \sin^2 \vartheta} \right] B_\varphi + \frac{\eta^{\prime}}{r} \frac{d(r B_\varphi)}{dr} \right\} &=& r \sin \vartheta {\bf B}_p \cdot \nabla \Omega,
\end{eqnarray}
where we exploited the fact that $\eta$ is a function of $r$ only, $\eta^{\prime} \equiv d \eta / d r$, and we introduced the following two "Reynolds numbers":
\begin{eqnarray*}
\mathscr{R}_\nu \equiv \frac{R_* u_A}{\nu_0} = \frac{t_\nu}{t_{\rm A}}, \\
\mathscr{R}_\eta  \equiv \frac{R_* u_A}{\eta_0} = \frac{t_\eta}{t_{\rm A}}.
\end{eqnarray*}

\subsection{Separation of variables}

To perform the separation of variables in the homogeneous PDEs associated to  Eqs. \eqref{eq:om_sc} and \eqref{eq:fl_sc}, we formally substitute factorised trial solutions:
\begin{eqnarray*}
\Omega(r,\vartheta,t) &=& \omega(t) Z(r,\vartheta), \mbox{~and~}\\
B_\varphi(r,\vartheta,t) &=& \beta(t) \Xi(r,\vartheta).
\end{eqnarray*}
The standard separation procedure leads to:
\begin{eqnarray}
\nonumber
\frac{d\omega}{dt} + \frac{\lambda}{\mathscr{R}_\nu} \omega &=& 0, \mbox{~and~} \\
\label{eq:saom}
\nabla \cdot ( \rho r^2 \sin^2 \vartheta \nu \nabla Z)  + \lambda \rho r^2 \sin^2 \vartheta Z &=& 0,
\end{eqnarray}
for Eq. \eqref{eq:om_sc}, introducing the separation constant $\lambda$; and:
\begin{eqnarray}
\nonumber
\frac{d\beta}{dt} + \frac{\mu}{\mathscr{R}_\eta} \beta &=& 0, \mbox{~and~} \\
\label{eq:safl}
\nabla \cdot \left( \eta \nabla \Xi \right) + \left[ \frac{\eta'}{r} - \frac{1}{r^2 \sin^2 \vartheta} \right] \Xi  + \mu \Xi &=&  0,
\end{eqnarray}
for Eq. \eqref{eq:fl_sc}, with $\mu$ being the separation constant.
Note that, since non-homogeneous source terms have been excluded for the moment,  the solution  asymptotically tends to $\Omega=$ const. and $B_\varphi=0$, and the separation constants $\lambda$ and $\mu$, appearing as eigenvalues in Eqs.\,\eqref{eq:saom} and \eqref{eq:safl}, have the physical meaning of reciprocal of the respective decay timescales.

The spatial, two-dimensional eigenvalue problems can be solved  by means of a further separation of variables, i.e., by assuming the following form for the $Z$ and $\Xi$ eigenfunctions:  
\begin{eqnarray*}
Z(r,\vartheta) &=& \zeta(r) P_{n}^{(1,1)}(\vartheta), \\
\Xi(r,\vartheta) &=& \xi(r) P_{n}^{1}(\vartheta),
\end{eqnarray*}
where the angular factors $P_{n}^{(1,1)}$ and $P_{n}^{1}$ are Jacobi polynomials and associated Legendre functions, respectively, which satisfy the following equations:
\begin{eqnarray}
\label{xx_eig}
\frac{d}{dx} \left[ (1-x^2)^2 \frac{d P_{n}^{(1,1)}}{dx}\right] + n(n+3)(1-x^2) P_{n}^{(1,1)} &=&0 , \\
\label{xx_eig2}
\frac{d}{dx} \left[ (1-x^2) \frac{dP_{n}^{1}}{dx}\right] + \left[ n(n+1) - \frac{1}{(1-x^2)}\right]  P_{n}^{1} &=&0,
\end{eqnarray}
where $x \equiv \cos \theta$, and $n$ is a non-negative integer. Such angular eigenfunctions are the only solutions of Eqs.~\eqref{xx_eig} and \eqref{xx_eig2} that are regular at the boundaries $x = \pm 1$, i.e. on the polar axis \citep[for Jacobi polynomials and associated Legendre functions see, e.g., ][]{Smirnov:1964, Abramowitz:1965book}.

After performing the separation of  variables, the eigenvalues of the angular eigenfunctions, i.e., $n(n+3)$ and $n(n+1)$, appear in the respective radial equations with the role of parameters. In other words, to each value of $n$ corresponds a radial eigenvalue problem for the  functions $\zeta_{nk}(r)$ and $\xi_{nk}(r)$, as defined by the equations:
\begin{eqnarray}
\label{rad_eigenf1}
\frac{1}{r^4}\frac{d}{dr}\left[ \rho \nu r^4 \frac{d\zeta_{nk}}{dr} \right] + \left[\lambda_{nk} \rho  - \frac{n(n+3)}{r^2} \rho \nu  \right]\zeta_{nk} &=& 0, \\
\label{rad_eigenf2}
\frac{1}{r^2}\frac{d}{dr}\left[ \eta r^2 \frac{d\xi_{nk}}{dr} \right] + \left[\mu_{nk}  + \frac{\eta'}{r} - \frac{n(n+1)}{r^2} \eta \right]\xi_{nk} &=& 0,
\end{eqnarray}
with the boundary conditions:
\begin{equation}
\begin{array}{cc}
\zeta_{nk}'(r_1) = \zeta_{nk}'(r_2) &= 0,\\
\xi_{nk}(r_1) = \xi_{nk}(r_2) &= 0,
\end{array}
\label{eq:zbc}
\end{equation}
where the primes denote derivation with respect to $r$. The radial eigenfunctions have therefore two indices: $n$ identifies the corresponding angular eigenfunction, while $k = 0, 1, \dots$ labels the different radial eigenvalues and eigenfunctions in the respective sets ($\lambda_{nk}$, $\zeta_{nk}$) and ($\mu_{nk}$, $\xi_{nk}$).

The radial equations \eqref{rad_eigenf1} and \eqref{rad_eigenf2} contain the functions $\rho(r)$, $\nu(r)$ and $\eta(r)$ tabulated from the stellar model and must  be solved numerically. The theory of St\"urm-Liouville problem ensures that the eigenvalues are real numbers, with a lower bound (i.e. a minimum eigenvalue that is  $\lambda_{n0}$ or $\mu_{n0}$) but no upper bound. The $k$-th eigenfunction, belonging to the $k$-th eigenvalue, is unique to within a multiplicative constant and has exactly $k$ zeros in the open interval $\ \left]r_1,r_2 \right[ \ $.

\subsection{Eigenfunction expansion}
\label{sec:expan}

For fixed $n$, the radial eigenfunction sets $\{ \zeta_{nk} \}$, $\{ \xi_{nk} \}$ are both orthogonal and complete in the interval $(r_1,r_2)$, in the sense that any piecewise continuous function of $r$ can be expanded as a generalised Fourier series of those functions in the same interval.
The same is true of $P^{(1,1)}_{n}$  and $P^{1}_{n}$ with respect to their index $n$ and for the angular variable $x \in [-1,1]$.

It is therefore possible to write the general solutions of Eqs. \eqref{eq:om_sc} and \eqref{eq:fl_sc} in the form:
\begin{align}
\nonumber
\Omega(r,\vartheta,t) & = \sum_{n,k} \omega_{nk}(t) \zeta_{nk}(r)P^{(1,1)}_n(\vartheta), \mbox{~and~}\\
\label{eq:solution}
\\
\nonumber
B_\varphi(r,\vartheta,t) &= \sum_{n,k} \beta_{nk}(t) \xi_{nk}(r)P^1_n(\vartheta),
\end{align}
respectively.
The generalised Fourier coefficients are functions of time and can be expressed through the scalar products defined by the following integrals:
\begin{eqnarray}
\label{eq:omcoeff}
\omega_{nk}(t) &\equiv&  \dfrac{\left< Z,\Omega \right>}{\left< Z,Z \right>} = \dfrac{ \displaystyle \int_{-1}^{+1} \int_{r_1}^{r_2} \rho r^2 \sin^2 \vartheta \zeta_{nk} P^{(1,1)}_n \Omega r^2 dr\ d\cos \vartheta }{\displaystyle \int_{-1}^{+1} \sin^2 \vartheta [P^{(1,1)}_n]^2 d\cos \vartheta  \int_{r_1}^{r_2} \rho r^2  [\zeta_{nk}]^2 r^2 dr} \mbox{~and~}\\
\nonumber
\\
\label{eq:flcoeff}
\beta_{nk}(t) &\equiv&  \dfrac{\left< \Xi,B_\varphi \right>}{\left< \Xi , \Xi \right>} = \dfrac{\displaystyle \int_{-1}^{+1} \int_{r_1}^{r_2} \xi_{nk}(r) P^{1}_n(\vartheta) B_\varphi(r,\vartheta,t) r^2 dr\ d\cos \vartheta }{\displaystyle \int_{-1}^{+1} [ P^{1}_n]^2 d\cos \vartheta \int_{r_1}^{r_2}  [\xi_{nk}]^2 r^2 dr},
\end{eqnarray}
where proper normalisation weights are introduced in the denominators, respectively. We recall that the radial eigenfunctions are orthogonal with respect to the weight function $w(r)= \rho r^{4} $ for Eq.~\eqref{rad_eigenf1} and $w(r) = r^{2}$ for Eq.~\eqref{rad_eigenf2}, respectively, appearing in the corresponding eigenvalue equations.
For simplicity of notation, in what follows we shall always assume that the eigenfunctions are normalised, i.e. $\left< Z_{nk},Z_{nk}\right> = 1$ and $\left< \Xi_{nk},\Xi_{nk}\right> = 1$.

\subsection{Complete time equations}
\label{complete_eqs}

The equations determining $\omega_{nk}(t)$ and $\beta_{nk}(t)$ as a function of $t$, including all the source terms in Eqs. \eqref{eq:om_sc} and \eqref{eq:fl_sc} and the non-homogeneous boundary conditions, are derived in Appendix \,\ref{app:selfadj} making use of the Green formula for the bi-dimensional self-adjoint operators in Eqs. \eqref{eq:saom} and \eqref{eq:safl}. They read:
\begin{equation}
\begin{array}{ccc}
\dfrac{d\omega_{nk}}{dt} + \dfrac{\lambda_{nk}}{\mathscr{R}_\nu} \omega_{nk}(t) = \displaystyle \sum_{mh} S_{nkmh}~\beta_{mh}(t)& + \dfrac{1}{\mathscr{R}_\nu} \mathscr{W}_{nk}(t), \\
& \\
\dfrac{d\beta_{nk}}{dt} + \dfrac{\mu_{nk}}{\mathscr{R}_\eta} \beta_{nk}(t) =\displaystyle \sum_{mh} T_{nkmh}~\omega_{mh}(t),&
\end{array}
\label{eq:time}
\end{equation}
This is a linear system of ODEs, containing $2\times N \times K$ equations, where $N$ and $K$ are the number of angular and radial eigenfunctions retained in the series in Eq. \eqref{eq:solution}, respectively. The system can be compactly rewritten as:
\begin{equation}
\label{eq:odecomp}
\dot{\bf x}(t) = \mathbb{A}\ {\bf x}(t) + \frac{1}{\mathscr{R}_\nu} {\bf w}(t),
\end{equation}
introducing the vector of the unknown functions, ${\bf x}(t) \equiv (\omega_{00}(t),\dots,\omega_{NK}(t),\beta_{00}(t),\dots,\beta_{NK}(t))$ and the matrix $\mathbb{A}$. Its properties are discussed in Appendix \ref{app:selfadj} where we show that it is a {\it normal matrix}, i.e.,  $\mathbb{A} \mathbb{A}^T = \mathbb{A}^T \mathbb{A}$. For a normal matrix, a generalised spectral theorem holds, which states that it is always possible to diagonalise it by means of a unitary transformation. The main difference with the well-known Hermitian case is that the diagonalised matrix is not necessarily real.
In other words, the matrix  $\mathbb{A}$ can be transformed into a diagonal matrix $ \Delta$ through a linear transformation of the kind:  $\Delta = \mathbb{U}^{-1} \mathbb{A} \mathbb{U}$, where $\mathbb{U}$ is a unitary matrix.  Through such a  transformation we reduce the solution of the set of Eqs. \eqref{eq:odecomp} to the integration of a decoupled set of ordinary differential equations of the kind:
\begin{equation}
\label{dec_eq}
\dot{y}_i(t) = \Delta_{ii} y_i(t) + g_i(t),
\end{equation}
with ${\bf y} = \mathbb{U}^{-1} {\bf x}$, ${\bf g}(t) = \mathbb{U}^{-1} {\bf w}(t)$, and $\Delta_{ii}$ the elements of the diagonal matrix $\Delta$. Such equations are immediately integrated as:
\begin{equation}
\label{ODE_sol}
y_{i} (t) = y_{i} (0) \exp (-\Delta_{ii} t) + \int_{0}^{t} g_{i}(t^{\prime}) \exp [-\Delta_{ii} (t-t^{\prime})] d t^{\prime}.
\end{equation}

Once the system \eqref{dec_eq} has been solved, we can go back to the original variables with the inverse transformation and therefore give the solution for the angular velocity and the toroidal field.

As already noted in Sect.~\ref{sec:method}, the non-homogeneous source terms containing the poloidal magnetic field produce a coupling between $\omega$'s and $\beta$'s, while the wind braking boundary condition has the effect of a driving force. Such a coupling allows the development of an oscillatory phase, despite the  parabolic character of each of Eqs. \eqref{eq:om_sc} and \eqref{eq:fl_sc} taken alone. From a mathematical point of view, the coupling terms give rise to the  antisymmetric part of $\mathbb{A}$ that is responsible for the appearance of the imaginary parts of the eigenfrequencies leading to such oscillations. On the other hand, the symmetric part of $\mathbb{A}$, which is diagonal, gives rise to the real part of the eigenfrequencies, which are negative, thus  producing a damping of the oscillations themselves. Therefore, the method introduced makes the evolution of torsional Alfv\'en waves more transparent from a mathematical point of view. Such oscillations are characterized by a periodic exchange of power between rotation and magnetic field, in the presence of a damping due to the diffusion of angular momentum and toroidal magnetic field (cf. Sect.~\ref{kin_energ}).

\subsection{Kinetic and magnetic energies}
\label{kin_energ}

The total rotational kinetic energy of the core is:
\begin{eqnarray}
\nonumber
T_\mathrm{kin} &=& \frac{1}{2} \int_\mathscr{C} \rho r^2 \sin^2 \vartheta \Omega^2(r,\vartheta,t) d^3 {\bf r}
= \frac{1}{2} \int_{r_1}^{r_2} \int_{-1}^{+1} \rho r^2 \sin ^2 \vartheta \sum_{n,k}\omega_{nk}\zeta_{nk}P^{(1,1)}_n \sum_{m,h}\omega_{mh}\zeta_{mh}P^{(1,1)}_m r^2 dr d\cos \vartheta  \\
&=& \frac{1}{2} \sum_{n,k} \sum_{m,h} \delta_{nm} \left( \int_{r_1}^{r_2} \rho r^2 \zeta_{nk} \zeta_{mh} r^2 dr \right) \omega_{nk} \omega_{mh}
= \frac{1}{2} \sum_{n,k} \sum_{m,h} \delta_{nm} \delta_{kh} \omega_{nk} \omega_{mh} = \frac{1}{2} \sum_{n,k} \omega_{nk}^2,
\label{eq:tkin}
\end{eqnarray}
where $\mathscr{C}$ is the volume of the core and we have exploited the orthogonality properties of the angular and radial eigenfunctions, respectively (the sum is intended over $n$, $k$, both going from $0$ to $\infty$).
Similarly, the total angular momentum of the core at any time can be written as:
\begin{eqnarray}
\nonumber
J_\mathrm{core}(t) &=&  \int_\mathscr{C} \rho r^2 \sin^2 \vartheta \Omega(r,\vartheta,t) d^3 {\bf r}  = \sum_{n,k} \left\{ \int_{-1}^{+1} \sin^2 \vartheta P^{(1,1)}_n P^{(1,1)}_0 d\cos\vartheta  \int_{r_1}^{r_2} \rho r^4 \zeta_{nk} \zeta_{00} dr \right\} \frac{\omega_{nk}(t)}{\zeta_{00} P^{(1,1)}_0} = \frac{ \omega_{00}(t) }{\zeta_{00} P^{(1,1)}_0},
\label{eq:jcore}
\end{eqnarray}
where the orthogonality of the eigenfunctions was exploited and also the fact that $\zeta_{00}=\rm{const.}$ and $P^{(1,1)}_0=\rm{const.}$
We see that the term involving $\omega_{00}$ in Eq. \eqref{eq:tkin} is proportional to the total angular momentum, thus it is always nonzero if rotation is present.
It is interesting to note that the state of minimum kinetic energy for a given total angular momentum corresponds to rigid rotation, with $T_\mathrm{kin}  = \frac{1}{2} \omega_{00}^2$.

In the same way as the kinetic energy, magnetic energy associated to the toroidal field component can be calculated as:
\begin{eqnarray}
U_B = \frac{1}{8\pi} \int_\mathscr{C} [B_\varphi(r,\vartheta,t)]^2 d^3 {\bf r} = \frac{1}{8\pi} \sum_{n,k} \beta_{nk}^2.
\label{eq:ubphi}
\end{eqnarray}

Note that Eqs. \eqref{eq:tkin} and \eqref{eq:ubphi} are closely related to the generalised Parseval relations for our Fourier series expansions \citep{Boyce:2001,Weinberger:1995}, that read:
\begin{eqnarray*}
\int_{-1}^{+1} \int_{r_1}^{r_2} \rho r^4 \sin^2 \vartheta \Omega^2(r,\vartheta,t) dr d \cos \vartheta &=& \sum_{n=0}^{\infty} \sum_{k=0}^\infty \omega_{nk}^2(t), \\
\int_{-1}^{+1}  \int_{r_1}^{r_2} r^2  B_\varphi^2(r,\vartheta,t) dr d \cos \vartheta &=& \sum_{n=0}^{\infty} \sum_{k=0}^\infty \beta_{nk}^2(t).
\end{eqnarray*}

The total energy contained in the system can be evaluated from $T_\mathrm{kin}$ and $U_B$, taking into account the proper scaling coming from nondimensionalisation, as:
\begin{equation}
\label{eq:etotal}
\frac{E_T}{E_0} = T_\mathrm{kin} + \left(\frac{t_\Omega}{t_A} \right)^2 U_B,
\end{equation}
where $E_0 = \frac{1}{2}I_0\Omega_0^2$ and $I_0=\rho_0 R_*^5$ are the dimensional scales of energy and moment of inertia, respectively.  

In an isolated and ideal (i.e., with negligible dissipation) MHD system, the total energy and angular momentum are conserved.
In a system with very high hydrodynamical and magnetic Reynolds numbers, energy conservation holds for a very long time, because dissipation is very small.
We shall make use of this property to apply energy conservation to our system, obtaining a relationship for the truncation error in Sect. \ref{sec:err_est}.
Note that there is a minimum energy state in an isolated dissipative system which is set by the conservation of the angular momentum. It corresponds to rigid rotation with $\Omega = \omega_{00} \zeta_{00} P^{(1,1)}_0$ and $B_\varphi=0$. This is the final state attained by our system, when dissipation of the kinetic energy of differential rotation and magnetic energy of the azimuthal field is over.

\subsection{Numerical issues}
\label{num_issues}

Our eigenfunction expansion technique
requires the numerical calculation of eigenvalues and eigenfunctions of two St\"urm-Liouville problems. This has been implemented using the shooting method described by, e.g., \citet{Press:1992}.  
For large values of $n$ or $k$, it is possible to resort to the asymptotic theory \citep{Morse_Feshbach:1953}.
The precision of these computations has been checked through the orthogonality of the eigenfunctions belonging to different eigenvalues, which is verified up to $10^{-5}$ in relative units in our numerical integrations.
One of the greatest advantages of our approach is that, once the determination of the set of eigenfunctions has been performed, the evolutionary calculations reduce to  a matrix diagonalisation and inversion, and the integration of a set of de-coupled first order, linear ODEs whose general solution is given by Eq.~\eqref{ODE_sol}.
Matrix inversion and diagonalisation are performed by means of the standard LAPACK routines\footnote{The routines are available at \texttt{http://www.netlib.org/lapack/}. } \citep{lapack}.

The possibility to estimate an upper bound for the error (see Sect. \ref{sec:err_est}) allows us to apply our method to compute benchmark solutions to test the accuracy of other methods to solve the coupled angular momentum and toroidal induction equations.
A remarkable advantage of our approach is that it allows us to calculate the solution at arbitrarily spaced intervals of time, without the need to match Courant-like criteria, which is usually a severe limitation for other numerical methods of solving parabolic PDEs.

The problem we are considering does not manifest the tendency to develop jump discontinuities, akin shocks in gas dynamics. This is another consequence of the parabolic nature of the equations we are solving, which tends to smooth out gradients with time. Therefore, our calculations are not significantly affected by the Gibbs phenomenon\footnote{The partial sum of a Fourier series prone to the Gibbs phenomenon systematically overshoots the true value of the function with spurious oscillations in the neighbourhood of any jump discontinuity. The problem is not alleviated by an increase of $N$ and $K$ because it is intrisic to the adopted representation of the solution \citep{Morse_Feshbach:1953}.} which can severely limit the quality of solutions computed by means of spectral methods.

Applying a spectral  method similar to ours to a problem already solved with different techniques, \citet{Cally:1991} showed that the accuracy of the solution does not degrade with time as far as the number of eigenfunctions included is greater than the number of modes actually excited.
With a resolution of $N=50$, $K=50$, we are able to properly model the evolution of our system when we adopt an enhanced diffusivity factor  $R_{\rm eff} = 10^{4}$, corresponding to a Reynolds number of the order of $10^{8}$ for the solar core (note, however, that our code works under the assumption of axisymmetry). For comparison, the best numerical MHD codes introduced so far, such as  ASH \citep{Clune:1999} which is also based on a spectral method, can reach Reynolds numbers of the order of $10^{5} - 10^{6}$ in fully 3D calculations \citep{Zahn_ea:2007}.  

\section{Results}
\label{sec:results}

To illustrate the results of our modelling it is convenient to define a reference case, constructed with representative values of the various parameters. Specifically, we assume $r_{1}=0.1\ R_{\odot}$, $r_{2}=0.7\ R_{\odot}$, with an initial rotation profile having $\Omega_{0}= 25 \ \Omega_\odot$, $\Delta \Omega/ \Omega_0=0.01$, $r_{\rm step}=0.5\ R_{\odot}$, $\Delta r_{\rm step}=0.2\ R_{\odot}$ (cf. Eq. \eqref{eq:omega_ic}).  
A confined dipole configuration (see Eq. \eqref{eq:dipstream}) is used for the poloidal field, with an amplitude $B_{0}=1$ G.

A result of previous works \citepalias{Charbonneau_MacGregor:1993,Rudiger_Kitchatinov:1996} is that molecular viscosity alone is not capable of reproducing the nearly rigid rotation of the radiative core within the solar age.  Therefore, we assume an effective viscosity enhancement factor $R_\mathrm{eff} = 10^4$, as defined in Eq. \eqref{eq:visc_enhance}. It suffices to reproduce a rigid rotation in the core at the age of the Sun  in our calculations, in agreement with helioseismic results.

For the reference model, we adopt $N=50$, $K=50$, and a grid of $1000$ points in the radial and meridional directions for the numerical evaluation of the radial and angular eigenfunctions and their scalar products, respectively.
The number of grid points was varied in the range $500 - 1000$ to check the consistency.
The current choice of $1000$ points was found to be sufficient for the quadrature routine to reach the accuracy of $\sim 10^{-5}$ in relative units.
The solution is calculated at $800$ time instants, logarithmically spaced  from the ZAMS to the present solar age \citep[for more details on the numerical implementation, see][]{Spada_PhD}.

Note that to obtain an accurate solution, it is necessary that the radial "eigenfunction resolution", i.e., $(r_2-r_1)/K$, be smaller than $r_1$.
With $r_1=0.1\ R_{\odot}$, and the choice $K \geq 30$, this condition is safely realised.

\subsection{Conservation of energy and estimate of the truncation error}
\label{sec:err_est}

Following Sect. \ref{kin_energ}, we define a global truncation error as:
\begin{equation*}
\sigma_{\rm NK}^2 \equiv E_T - \sum_{n=0}^N \sum_{k=0}^K \left\{ \omega_{nk}^2 + \left(\frac{t_\Omega}{t_A} \right)^2\beta_{nk}^2 \right\} = \sum_{n=N}^\infty \sum_{k=K}^\infty \left\{ \omega_{nk}^2 + \left(\frac{t_\Omega}{t_A} \right)^2\beta_{nk}^2 \right\}.
\end{equation*}
From the last equality, we see that $\sigma_{NK}^2$ is simply the sum of the squares of the coefficients of the neglected terms, and is thus always positive.
\begin{figure}
\begin{center}
\includegraphics[width=10truecm,angle=90]{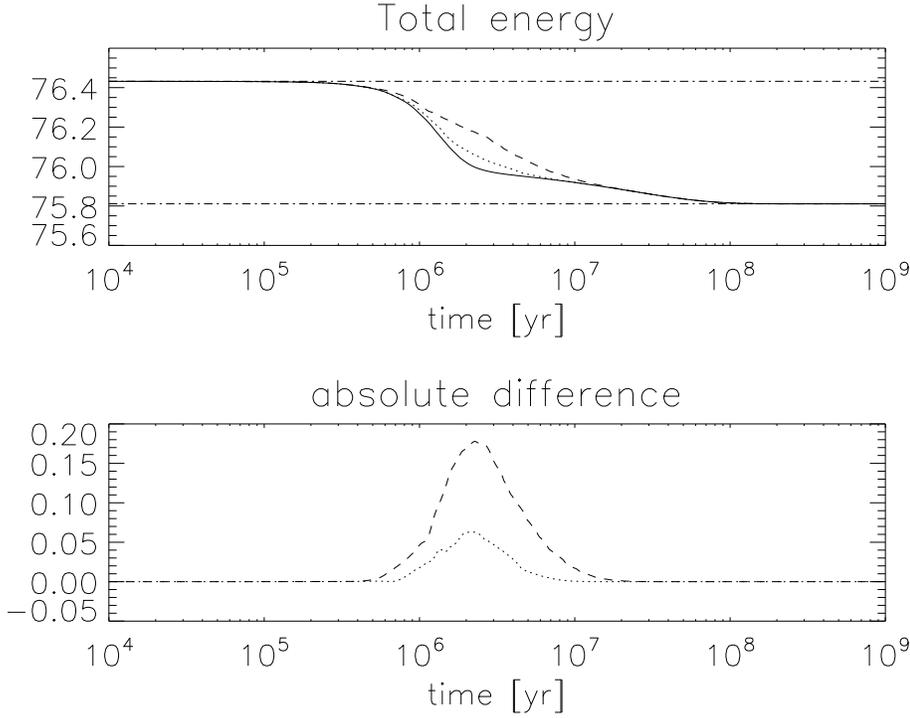}
\caption{Upper panel: Evolution of the model without wind braking for resolutions $N\times K = 20 \times 20$ (dashed line), $30 \times 30$ (dotted line), and $50 \times 50$ (solid line), respectively. The extreme values of the energy $E_{T}^\mathrm{in}$ and $E_T^\mathrm{min}$, as given by Eqs. \eqref{eq:Ein} and \eqref{eq:Emin}, are also shown (dash-dotted lines). Lower panel: absolute difference between the reference resolution ($50 \times 50$) and the $20 \times 20$ (dashed line) and $30 \times 30$ (dotted line) solutions.}
\label{fig:error}
\end{center}
\end{figure}

Test calculations were performed without wind braking, an initial differential rotation $\Delta \Omega/\Omega_0 = 0.25$, and $N=K=20, \ 30$ and $50$, respectively.
This suffices to test the code performance at different resolutions (i.e. with different values of $N$ and $K$), focusing on the role of the coefficients corresponding to the smallest spatial scales.
Testing the complete solution, namely including wind braking, is in principle feasible, but this would essentially influence only the $\omega_{00}$ term.

As discussed in Sect. \ref{kin_energ}, in our high Reynolds number regime there is an initial phase, lasting $\sim 3\cdot 10^5$ yr, when total energy is conserved to a high degree (see Fig. \ref{fig:error}). Then dissipation begins to decrease the energy down to the minimum value set by angular momentum conservation.  A complete dissipation takes place within $10^8$ yr, independently of the adopted spatial resolution.

Concerning the estimate of the truncation errors, all the calculations exhibit an excellent agreement (with $\sigma_{\rm  NK} \sim 10^{-4}$ in relative units) with the exact values of the initial and final energies $E_T^\mathrm{in}$ and $E_T^\mathrm{min}$ during the initial and final stages of the evolution, respectively. They can be computed from:
\begin{equation}
E_T^\mathrm{in} \equiv \frac{2}{3} \int_{r_1}^{r_2} \rho r^4 \left[ \frac{\Omega_\mathrm{in}(r)}{\Omega_0}\right]^2 dr
\label{eq:Ein}
\end{equation}
and
\begin{equation}
E_T^\mathrm{min} \equiv \frac{1}{2} \frac{J^2_\mathrm{core}(0)}{I_\mathrm{core}},
\label{eq:Emin}
\end{equation}
with $I_\mathrm{core}$ the total moment of inertia of the core scaled to $I_0$, i.e.:
\begin{equation*}
I_\mathrm{core} = \frac{4}{3} \int_{r_1}^{r_2}\rho r^4 dr.
\end{equation*}

The comparison of the less resolved runs with the reference run, $50 \times 50$, is particularly interesting during the intermediate phase, i.e. between $\sim 10^6 - 10^7$ yr, when the largest differences appear (see, in particular, the lower panel of Fig. \ref{fig:error}). Apparently, our reference solution reproduces more accurately the dissipation rate, given by the slope of the corresponding plot in the upper panel of Fig. \ref{fig:error}, owing to the inclusion of a larger number of terms in the series. After about $2/3$ of the total energy dissipation has taken place, a change of the slope occurs also in the reference solution. If this is a numerical artefact due to the truncation of the energy transfer to smaller spatial scales, then our computed time-scales for relaxation to a rigid rotation state could be slightly overestimated. However, this is not a serious limitation if we are interested in an order of magnitude estimate of the coupling time-scale akin to, for example, the parameter $\tau_c$ in the model by \citet{Allain:1998} \citep[see also][]{Bouvier:2008b}.
A suitable correction factor could, in principle, be estimated from Fig. \ref{fig:error}, by extrapolating the slope of the energy variation vs. time beyond $\sim 2 \cdot 10^6$ yr. With this simple approach, we can estimate that the e-folding time for energy dissipation, as derived from the upper part of the plot of the energy vs. time, is overestimated by no more than a factor of $2-3$.

\subsection{Evolution of the reference model}

The model with wind braking evolves through the following three main phases:

\begin{itemize}
\item[a)] \emph{Linear build up of toroidal field}

The toroidal field amplitude, starting from zero with the initial condition in Eq. \eqref{eq:bphi_ic}, grows with time for a few $t_{\rm \Omega}$, with  $t_{\Omega} \ll t_{\rm A} \sim 3.5\cdot 10^3$ yr for $B_{0}=1$ G.
In this phase, advection dominates over diffusion in the induction equation \eqref{eq:fl_sc}, so that:
\begin{equation*}
\frac{\partial B_\varphi}{\partial t} \simeq r \sin \vartheta {\bf B}_p \cdot \nabla \Omega.
\end{equation*}
Noting that, in the initial phases, $\nabla \Omega = \frac{\partial \Omega}{\partial r} \bf r$ (cf. Eq. \eqref{eq:omega_ic}) and integrating with respect to the time:
\begin{equation}
\label{eq:intor}
B_\varphi(r,\vartheta,t) \simeq  \ r \sin \vartheta B_r  \frac{\partial \Omega}{\partial r} \Delta t.
\end{equation}
The stretching of the poloidal field lines by differential rotation is characterised by the smallest time-scale, namely $t_\Omega$. The amplitude of $B_\varphi$  increases linearly with time, feeding on the shearing of ${\bf B}_p$, and is proportional to the initial differential rotation $\Delta \Omega$ and poloidal field amplitude $B_0$.
\begin{figure}
\begin{center}
\includegraphics[width=14truecm]{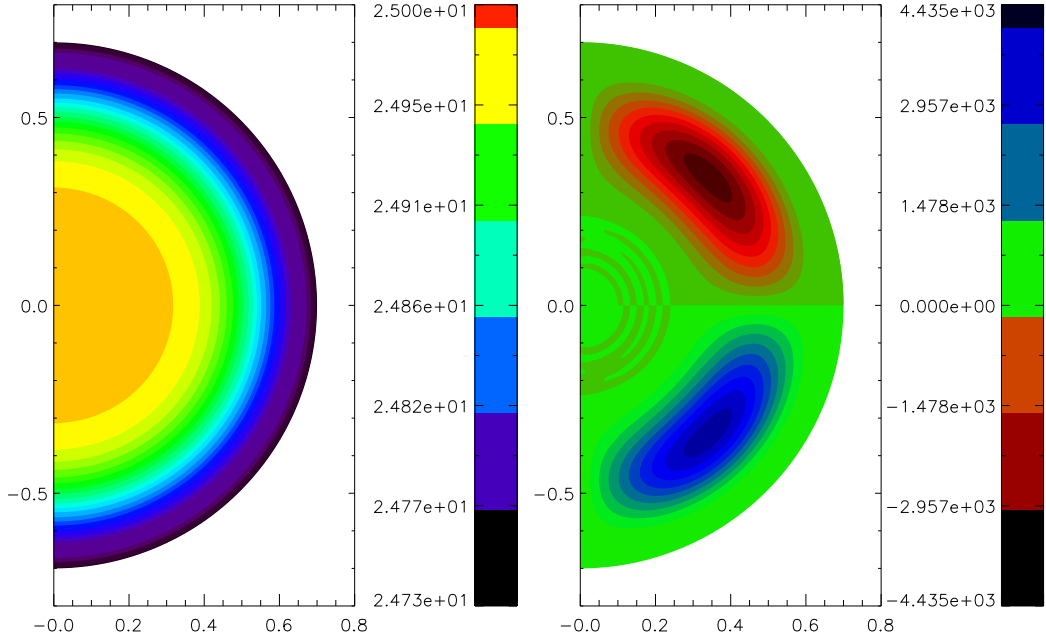}
\caption{Isocontour plots of $\Omega$ (left panel, in units of $\Omega_\odot$) and $B_\varphi$ (right panel, in $G$) after  $10^{-3} t_{\rm A}$.}
\label{fig:contlin}
\end{center}
\end{figure}
The  toroidal field is antisymmetric with respect to the equator, as shown in the contour plot of Fig. \ref{fig:contlin}.
This is a  consequence of Eq. \eqref{eq:intor}, since $\Omega$ is initially independent of $\vartheta$ and $B_r \sim \sin \vartheta \cos \vartheta$, as can be seen from Eq. \eqref{eq:dipstream} for $\Psi$.

The linear growth of the field eventually ends when a balance is established between the torques due to wind braking and the azimuthal component of the Lorenz force, which reacts to further shearing.  

\item[b)] \emph{Torsional Alfv\'en waves}
\label{torsional_waves}

The second phase, ranging approximately from $10^3$ to $10^6$ yr for $B_{0}=1$ G, is characterised by the excitation and progressive damping of torsional Alfv\'en waves.
When the Lorentz force becomes effective, it acts as a restoring force on any further stretching of the poloidal field lines.
This excites oscillations of $B_\varphi$ and $\Omega$, propagating along each poloidal field line (cf. Sect. \ref{complete_eqs} and Fig. \ref{fig:contosc}).
These waves have a linear behaviour for arbitrary amplitude, because their phase velocity, $V_{\rm Ap}={{ B}_p}/{\sqrt{4 \pi \rho}}$, is independent of the toroidal field intensity or the rotation rate.
\begin{figure}
\begin{center}
\includegraphics[width=14truecm]{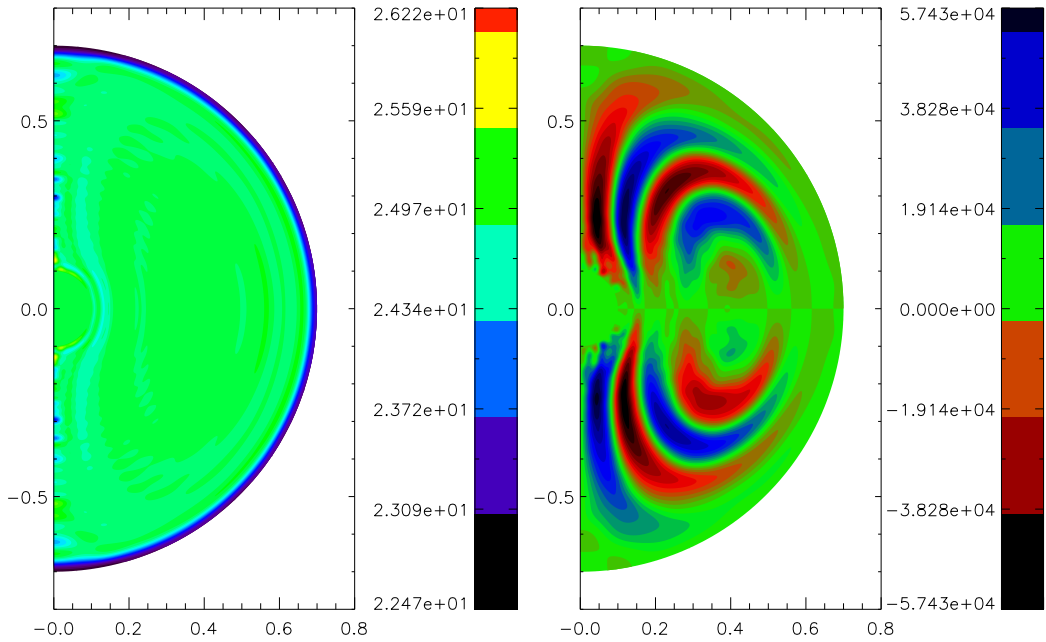}
\caption{As in Fig. \ref{fig:contlin}, for $t=3.1\cdot 10^3 $ yr $ \simeq  1\ t_A$. Out-of-phase oscillations on neighbour poloidal field isosurfaces are clearly evident in the $B_\varphi$ isocontours in the right panel.}
\label{fig:contosc}
\end{center}
\end{figure}
Neighbouring poloidal field lines, however, oscillate with slightly different phase velocities because $|{\bf B}_p|$ and $\rho$ are not uniform in the core. This leads to the development and progressive increase of a phase lag between waves propagating along neighbour field lines.  Eventually these waves get in opposition of phase and dissipate quickly. This process is called \emph{phase mixing} \citep{Spruit:1987,Cally:1991,Spruit:1999}.
\begin{figure}
 \begin{center}
   \subfigure[$B_\varphi$ at fixed latitude.]{\label{fig:osc-a}
\includegraphics[scale=0.25,angle=90]{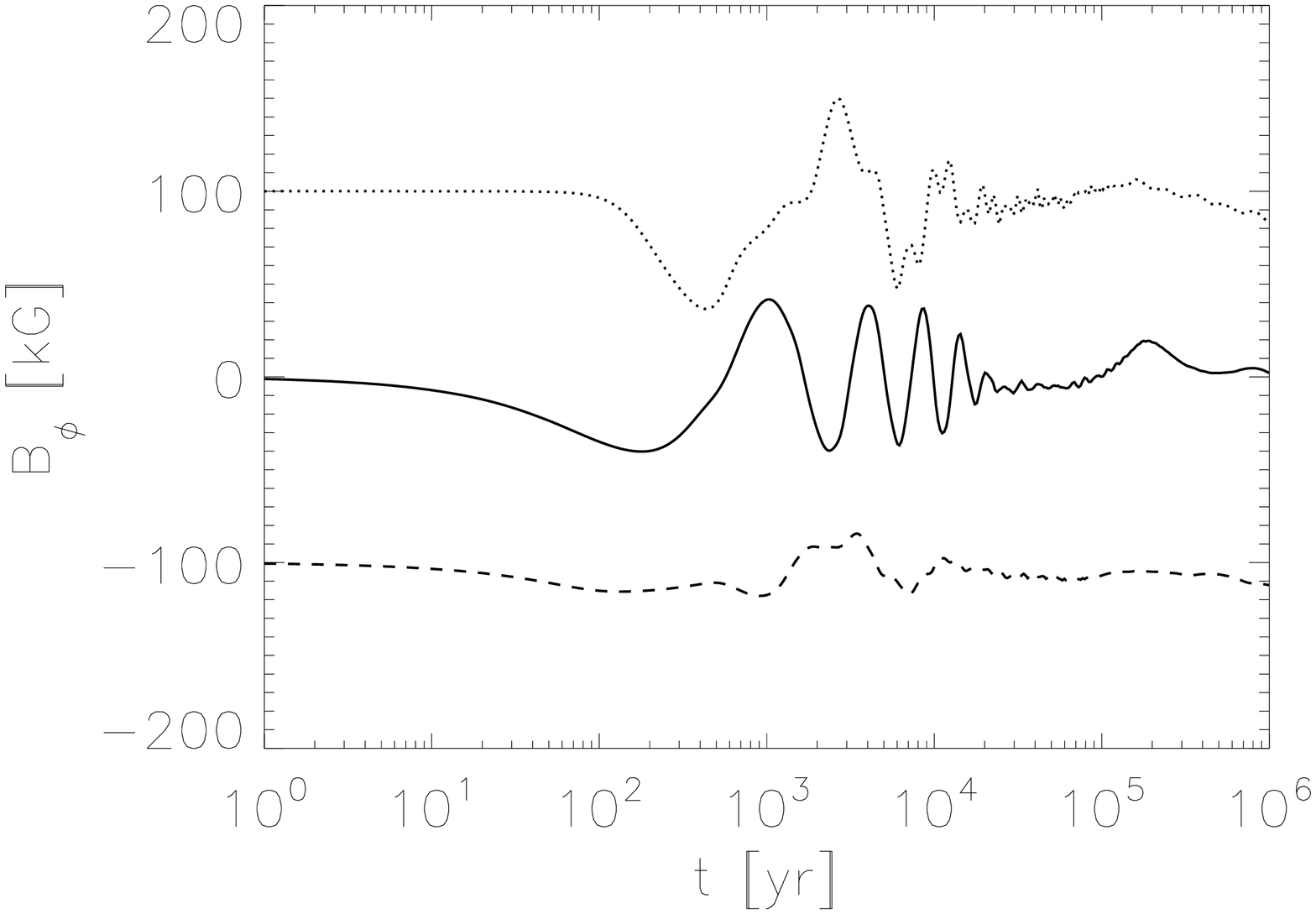}}
   \subfigure[$B_\varphi$ at fixed radius.]{\label{fig:osc-b}
\includegraphics[scale=0.25,angle=90]{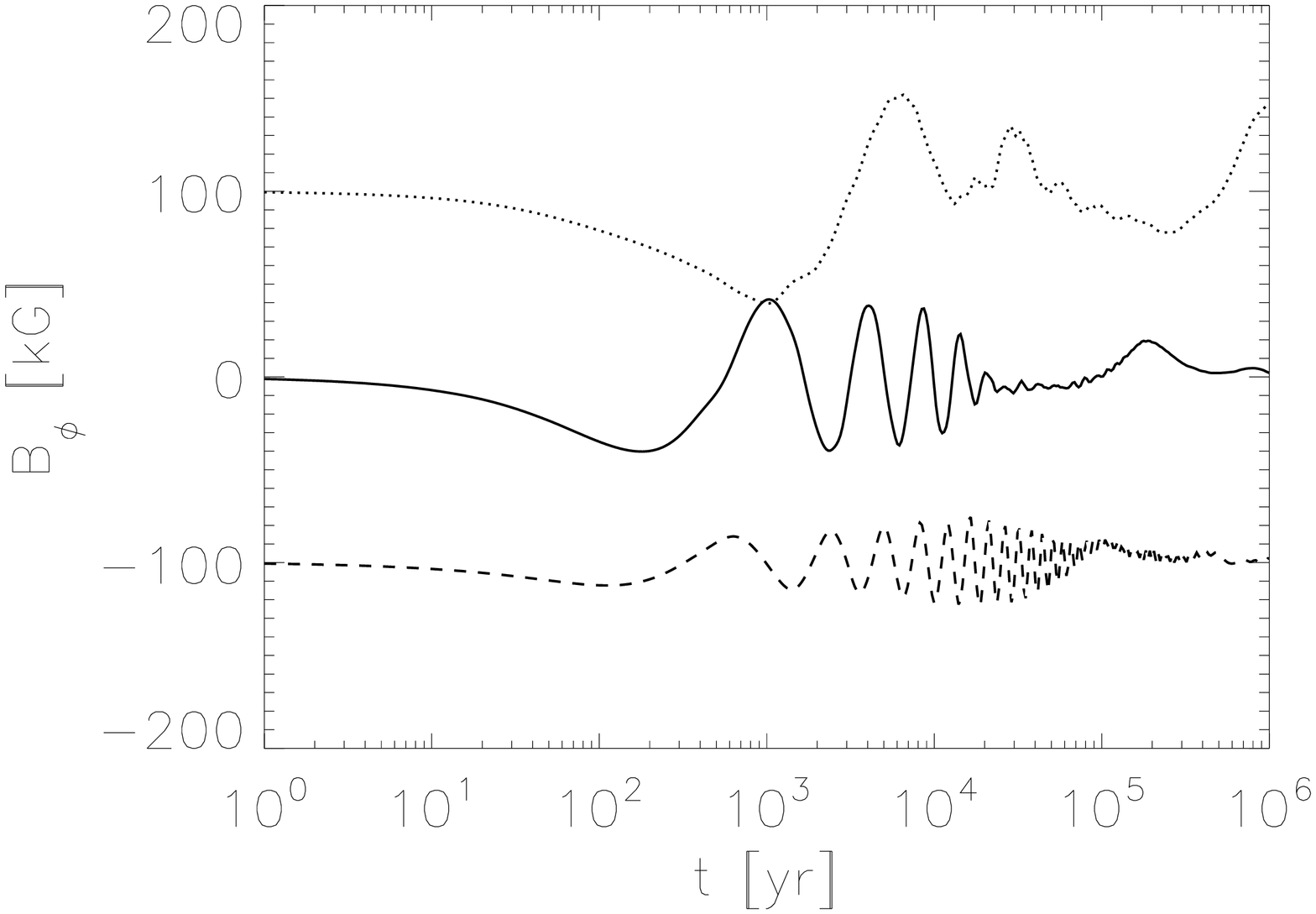}} \\
   \subfigure[$\Omega$ at fixed latitude.]{\label{fig:osc-c}
 \includegraphics[scale=0.25,angle=90]{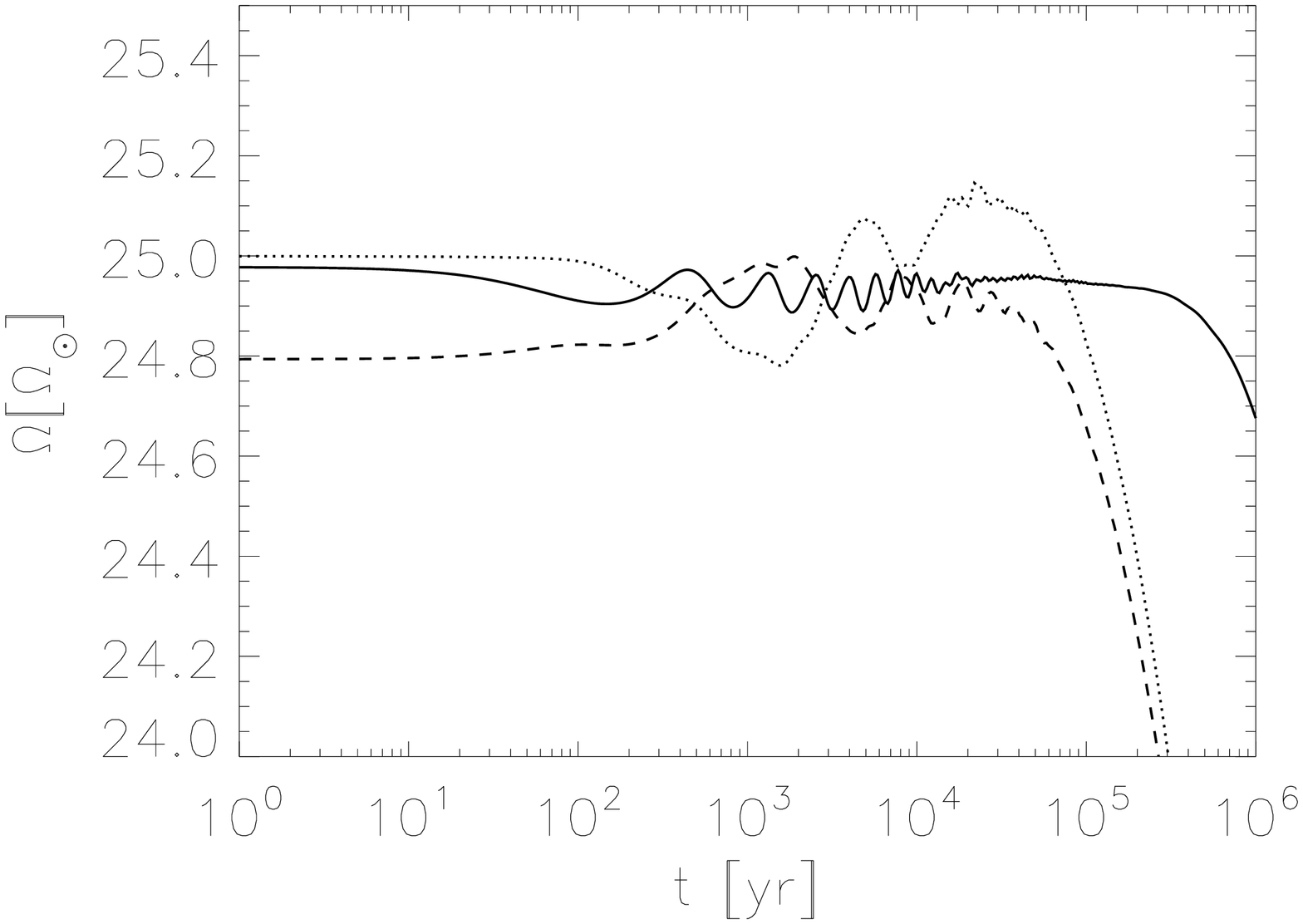}}
       \subfigure[$\Omega$ at fixed radius.]{\label{fig:osc-d}  \includegraphics[scale=0.25,angle=90]{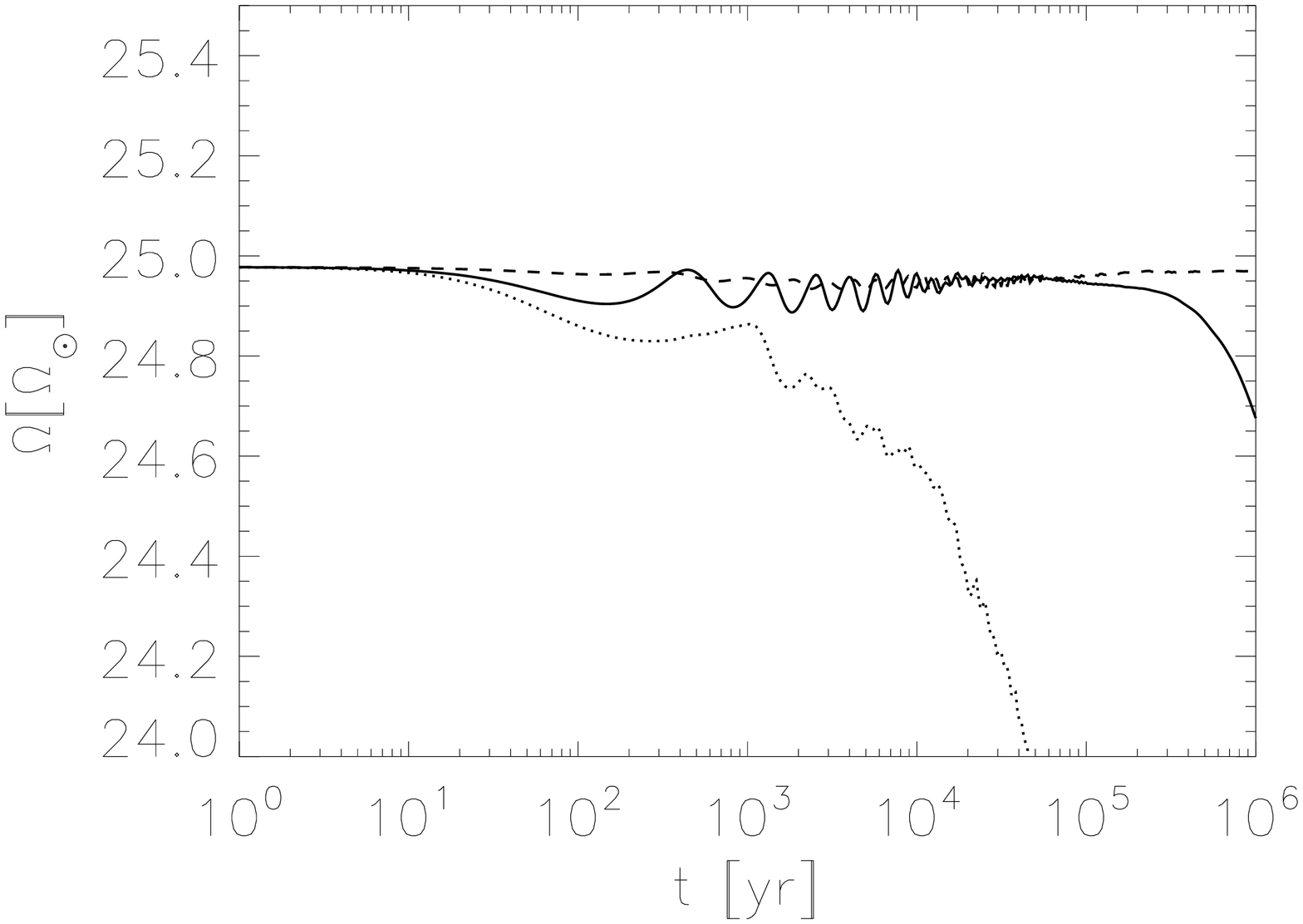}}
 \end{center}
 \caption{$B_\varphi$ (upper panels) and $\Omega$ (lower panels) vs. time at some control points. On the left panels, points at $r=0.2, \ 0.4, \ 0.6 \; R_{\odot}$, with  $\vartheta=\pi/4$ fixed, are indicated by dotted, solid and dashed lines, respectively. On the right panels, points at  $\vartheta=\pi/12, \ \pi/4, \ 5\pi/12$, with $r=0.4 \; R_{\odot}$ fixed,  are indicated by  dotted, solid and dashed lines, respectively. Dotted and dashed curves in \ref{fig:osc-a} and \ref{fig:osc-b} were given an offset of $\pm 100$ kG for the sake of clarity.}
 \label{fig:osc}
\end{figure}

Some interesting features of the oscillations are illustrated in Fig. \ref{fig:osc}.
Control points located at different depths show that oscillations in the field amplitude are more readily excited and damped in the external part of the core, i.e. closer to the driving torque (Fig. \ref{fig:osc-a}).
The differential rotation imposed as an initial condition is significantly smoothed out before the wind braking becomes important (Fig. \ref{fig:osc-c}).
Control points located at the same radius were chosen to investigate the effect of a trapping of the oscillations inside the so-called ``dead zone" of the poloidal field, i.e. in the domain near the magnetic neutral "O" point at $r\sim 0.4\ R_{\odot}$ on the equator.  
At lower latitudes, the $B_\varphi$ oscillations persist for a longer time (Fig. \ref{fig:osc-b}) and rotational braking is delayed (Fig. \ref{fig:osc-d}). We conclude that,  
with this choice of $R_\mathrm{eff}$, oscillations are already damped in a considerable fraction of the core, including the dead zone, before the wind braking of the star as a whole becomes important.  

Our analytic treatment allows us to evaluate precise upper and lower bounds for the total energy contained in the subgrid lenghtscales during the phase mixing process. For instance, considering the
cases plotted in Fig.~\ref{fig:error} and that the total energy is always greater than the sum of the squared amplitudes of the included modes, we find that $2.4 \times 10^{-3} \leq \sigma^{2}_{NK}(t)/E_{\rm T}(t) < 9.1 \times 10^{-3}$ at any time $t$ during the phase mixing evolution. This rigorous treatment of the subgrid contribution is a new and interesting feature of our analytic approach.

\item[c)] \emph{Quasi-stationary evolution}

For the rest of the computation, the evolution proceeds at a slower pace, in a quasi-stationary regime reminiscent of Ferraro isorotation, i.e., with angular velocity almost uniform along each poloidal field line. Actually, this is the expected outcome of the phase-mixing of torsional waves, which smooths out the fluctuations of angular velocity on poloidal field isosurfaces, that is the surfaces where $\Psi = \mathrm{const}$. Neglecting the diffusion processes, the stationary condition is reached when:
\begin{eqnarray}
\begin{array}{cc}
{\bf B}_p \cdot \nabla \Omega & = 0 \\
{\bf B}_p \cdot \nabla (r \sin \vartheta B_\varphi) & = 0.
\end{array}
\label{eq:qst}
\end{eqnarray}
Eqs. \eqref{eq:qst} predict that the rotation rate must be constant on $\Psi$ isosurfaces, and that the azimuthal component of the Lorentz force must be zero \citep{Garaud_Guervilly:2008}.
These conditions are achieved at late times in our model, as can be seen in Fig. \ref{fig:contqs}.
This is a consequence of the ratio of the time-scales involved, because deviations from the conditions in Eqs. \eqref{eq:qst} are produced on the wind braking or diffusive time-scale, i.e., from $\sim 10^6$ to $\sim 10^9$ yr, but are compensated for on the much shorter Alfv\'en time-scale $t_{\rm A} \sim 10^3$ yr \citepalias[cf. ][]{Charbonneau_MacGregor:1993}.
Nevertheless, the magnetic (and viscous) stresses are still acting to transfer angular momentum on the long time-scale characteristic of the late wind braking process.
\begin{figure}
\begin{center}
\includegraphics[width=14truecm]{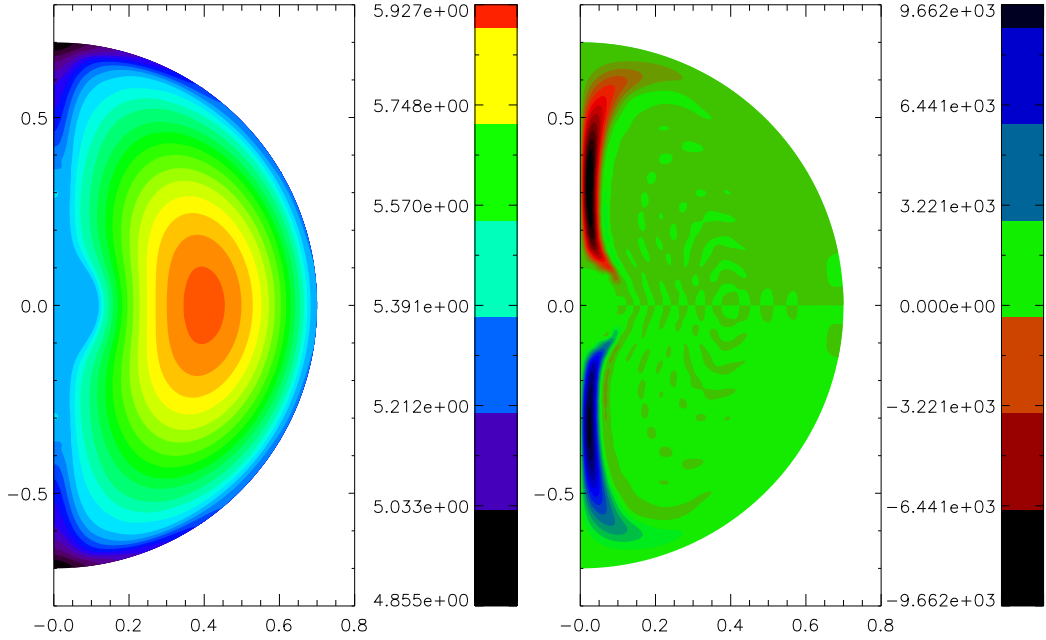}
\caption{As in Fig. \ref{fig:contlin}, for  $t=1.23$ Gyr or $ 350\,000\ t_A$. $\Omega$ (left panel) and $B_\varphi$ (right panel) isolines do not change for the rest of the evolution, apart from an overall scale factor. Note that $\Omega$ has become nearly constant on poloidal field lines and $B_\varphi=0$ inside the ``dead zone'', where $r \sin \vartheta B_\varphi$ must be both antisymmetric with respect to the equator and constant along field lines.}
\label{fig:contqs}
\end{center}
\end{figure}

As already noted, a model with $R_\mathrm{eff}=1$ does not attain a condition of uniform rotation in the core within the solar age, even in the presence of a large scale poloidal field.  Magnetic transport of angular momentum due to phase mixing is effective only along poloidal field lines, while the only means to couple the plasma across magnetic surfaces is by the action of the viscosity.
Molecular viscosity alone fails to ensure an effective rotational coupling within the dead zone of  the poloidal field because its coupling time-scale is of the order of $R^{2}/\nu$, i.e., longer than the age of the Sun. In the model considered here, it is an enhanced viscosity which eventually enforces rigid rotation, as seen in Fig. \ref{fig:glevrf}.
\begin{figure}
\begin{center}
\includegraphics[width=8truecm,angle=90]{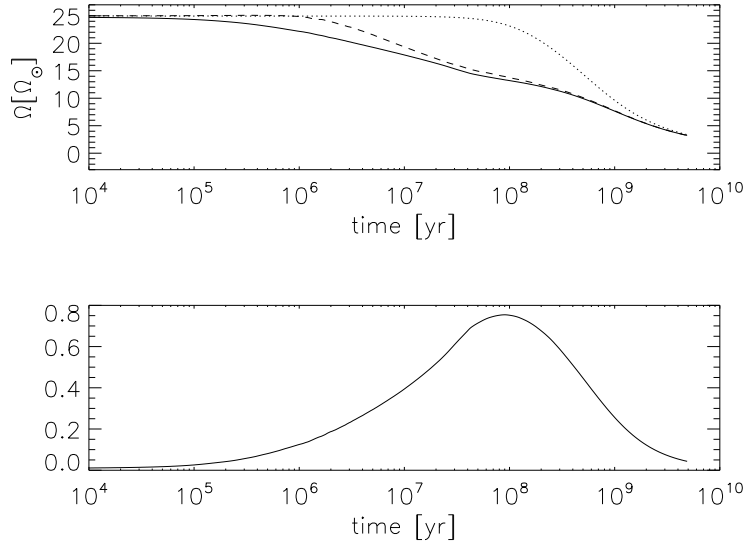}
\caption{Upper panel: latitudinal average of the angular velocity vs. time at three different depths: $r_2$ (solid line), $r_1$ (dashed line), and $r_c=0.4 \; R_{\odot}$ (dotted line), in the vicinity of the neutral magnetic point. Lower panel: relative difference between the average angular velocities at $r_2$ and $r_c$ vs. time.}
\label{fig:glevrf}
\end{center}
\end{figure}

Note that the evolution of the average angular velocity is very similar to that obtained by \citetalias{Rudiger_Kitchatinov:1996} \citepalias[see the upper panel of Fig. \ref{fig:glevrf} and Fig. 4 of][]{Rudiger_Kitchatinov:1996}, apart from a shallow dip in our curve around $10^6$ yr due to the inclusion of a saturation for the angular momentum loss in the wind braking law adopted in our model.

The two panels in Fig. \ref{fig:omprof} show plots of the angular velocity as a function of the fractional radius for two representative latitudes, corresponding to viscosity enhancements $R_{\rm eff}=10$ and $R_{\rm eff}=10^4$, respectively.
For comparison purposes, they are presented with the same axes of the upper panel in Fig. 6 and the central panel in Fig. 7 of \citetalias{Rudiger_Kitchatinov:1996}, respectively.
The profile near the pole is not shown here because the exclusion of the central part of the core ($r<r_1$) produces spurious oscillations near the axis that are not damped effectively with a viscosity value as low as $R_{\rm eff}$.
Given the negligible amount of angular momentum near the axis, however, this does not influence the whole calculation significantly.

The error on the angular velocity was estimated for both cases according to the method of Sect. \ref{sec:err_est}, obtaining a relative error $ \lesssim 4 \, \%$ for $R_{\rm eff}=10$ (corresponding to a Reynolds number $Re \sim 10^{11}$) and $ \lesssim 2.5 \, \%$ for $R_{\rm eff}=10$ ($Re \sim 10^{8}$).
The appropriate error bars are shown in Fig. \ref{fig:omprof}.

The small oscillations in the left panel of Fig. \ref{fig:omprof} are well within the error bars and are a consequence of the difficulty to represent a nearly constant plateau with a finite number of periodic eigenfunctions.
Overall, both panels compare well with their counterparts in \citetalias{Rudiger_Kitchatinov:1996}, apart from the immediate vicinities of the radial boundaries $r_1$ and $r_2$, where the effect of the different boundary conditions is more marked.
The stronger coupling between the equator and $45^\circ$ that can be seen in our results is a consequence of the different prescription for the poloidal field flux function.


If conditions close to Eqs. \eqref{eq:qst} are eventually established, a poloidal field having a multipole configuration of an order higher than dipolar could be more efficient in imposing rigid rotation in the whole stellar core because its dead zones are significantly smaller \citep{Spruit:1999}. This is in fact observed in the case of our quadrupolar model, that attains a solid body rotation within $\sim 1$ Gyr (see Sect. \ref{sec:DRvisc}). Such a time-scale is to be compared with a decay time of the quadrupolar mode of $2.7$ Gyr, as estimated by \citetalias{Rudiger_Kitchatinov:1996}.
\end{itemize}

\begin{figure}
 \begin{center}
   \subfigure
 {\label{fig:omprof-a}
\includegraphics[scale=0.45]{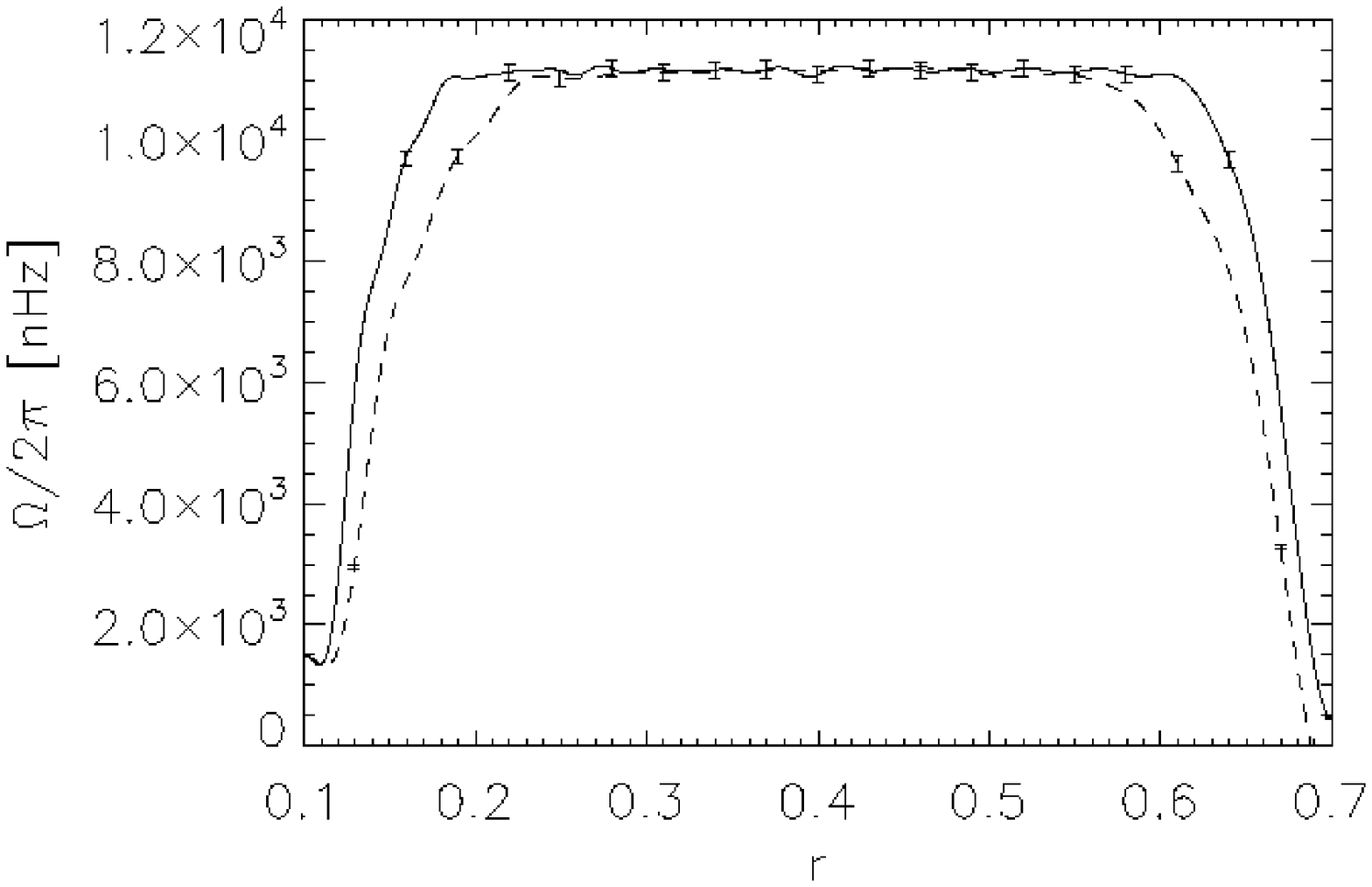}}
   \subfigure
   {\label{fig:omprof-b}
\includegraphics[scale=0.45]{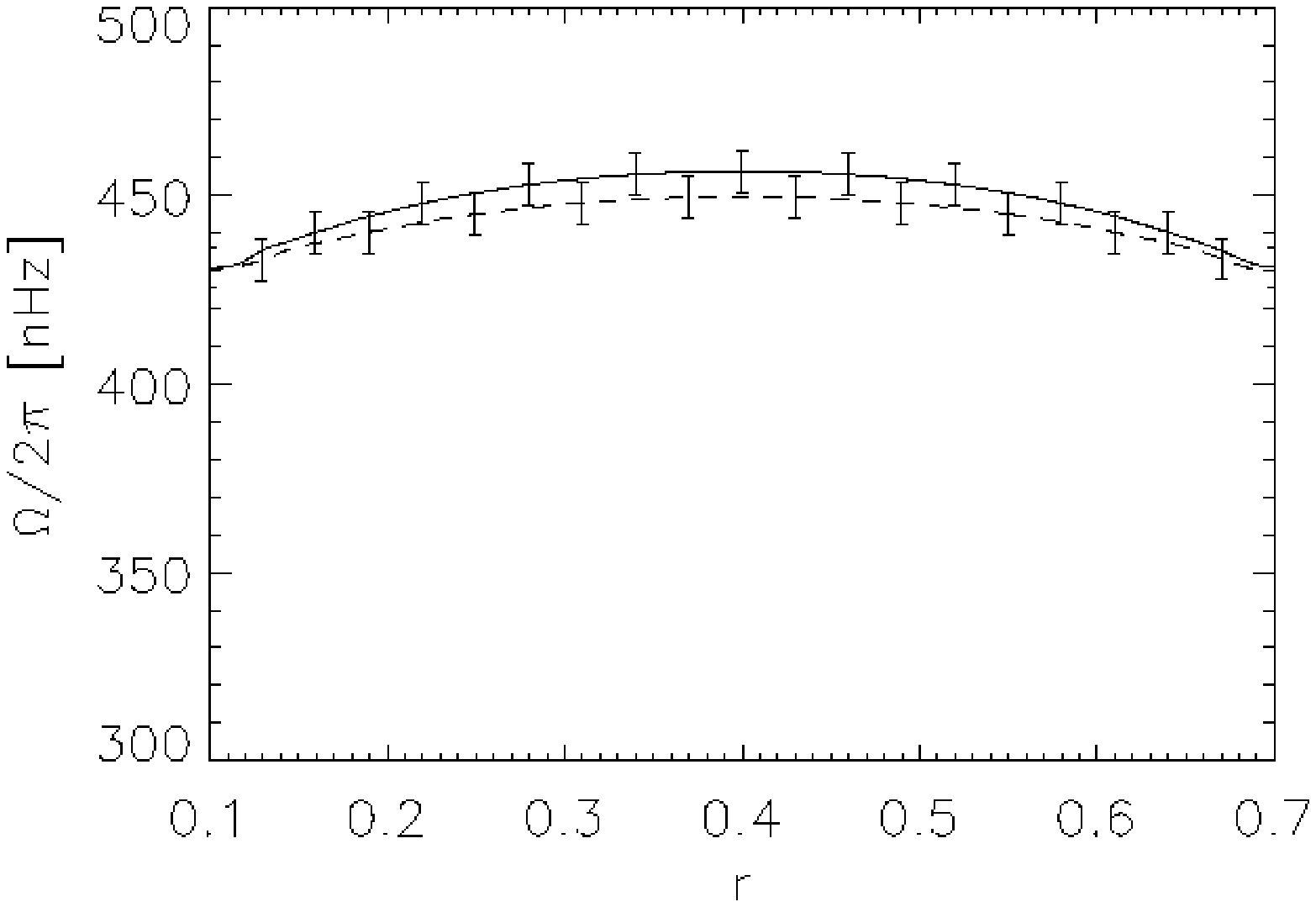}} 
  \end{center}
 \caption{Angular velocity as a function of the fractional radius $r$, at the equator (solid line) and at $45^\circ$ latitude (dashed line), at the age of the Sun ($\sim 4.6$ Gyr), for $R_{\rm eff}=10$ (left panel) and $R_{\rm eff}=10^4$ (right panel).}
 \label{fig:omprof}
\end{figure}

\subsection{Differential rotation of the core}
\label{sec:DRvisc}

To ease the comparison with previous works, a measure of core differential rotation is defined as:
\begin{equation}
\label{eq:cm_dec}
D(t) \equiv  \frac{3}{2(r_2^3-r_1^3)} \int_{-1}^{+1} \int_{r_1}^{r_2} \frac{\Omega(r,\vartheta,t)-\Omega(r_2,\pi/2,t)}{\Omega(r_2,\pi/2,t)} dr \, d\cos \vartheta.
\end{equation}
\begin{figure}
\begin{center}
\includegraphics[width=8truecm,angle=90]{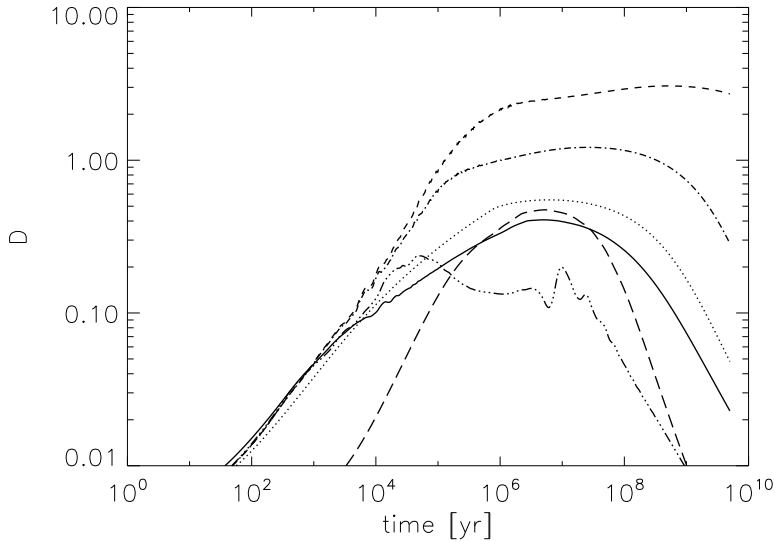}
\caption{
Differential rotation measure $D$ vs. time for different models. Solid line: reference model (dipolar poloidal field with $B_{0}=1$~G and $R_\mathrm{eff}=10^4$); dashed line: model with $R_\mathrm{eff}=10^2$; dash-dotted line: model with $R_\mathrm{eff}=10^3$; long-dashed line: model with $R_\mathrm{eff}=10^5$; dash-triple-dotted line: model with quadrupolar poloidal field ($B_{0}=1$ G) and $R_\mathrm{eff}=10^4$; dotted line: non-magnetic model (i.e., $B_{0}=0$) with $R_\mathrm{eff}=10^4$. The oscillations of $D$ near $10^4$ and $10^7$ yr are due to  torsional Alfv\'en waves.}
\label{fig:decb}
\end{center}
\end{figure}

In Fig.~\ref{fig:decb} we show the evolution of $D(t)$ for some models compared with the reference model.

The curves corresponding to $R_\mathrm{eff} = 10^2 - 10^3$ show that such low values of the viscosity are not enough to attain a uniform rotation within the age of the Sun.
Notably, decreasing $R_\mathrm{eff}$ has the effect of attaining the maximum level of internal decoupling at later times.

\begin{figure}
\begin{center}
\includegraphics[width=8truecm,angle=90]{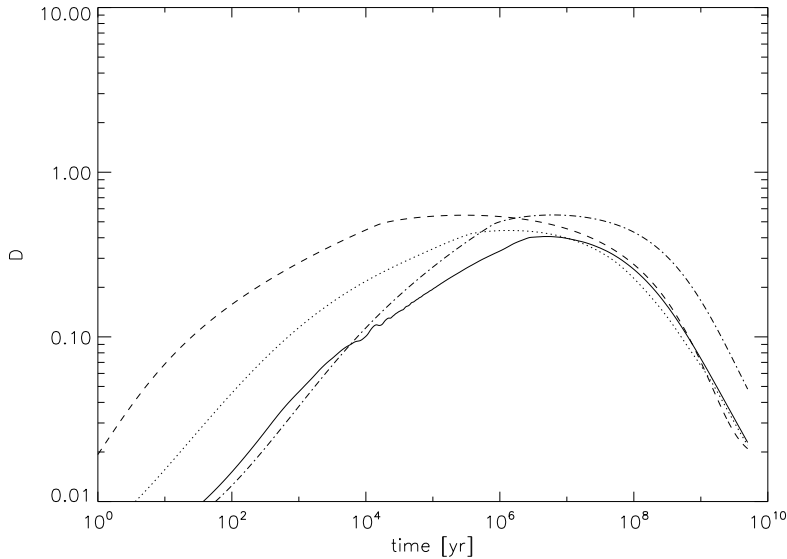}
\caption{
$D$ vs. time for different poloidal field intensity. Solid line: reference model ($B_0=1$ G); dotted: $B_0=0.1$ G; dashed: $B_0=0.01$ G; dash-dotted: non-magnetic model. The oscillations of $D$ near $10^4$ and $10^7$ yr are due to  torsional Alfv\'en waves.}
\label{fig:varb0}
\end{center}
\end{figure}

The poloidal field geometry has a remarkable impact on the phase-mixing phase and on the quasi-stationary regime that is subsequently established.
Apart from the changes in the shape of the angular velocity and toroidal field isolines, an assigned flux function with quadrupolar symmetry determines both qualitative and quantitative modifications in the differential rotation evolution with respect to that with a dipolar symmetry.
As it is shown in Fig. \ref{fig:decb}, the decoupling phase is shortened and $D(t)$ presents a {plateau} around $10^6 - 10^7$ yr in the quadrupolar model. Conversely,  
the role played by the poloidal field intensity $B_{0}$ is quite modest, as is apparent in Fig. \ref{fig:varb0}, because the winding up by differential rotation leads in any case to similar azimuthal  stresses $B_{r} B_{\varphi}$ at the end of the linear build up phase of the azimuthal field.

Our reference model behaves very similarly to that computed by \citetalias{Charbonneau_MacGregor:1993} with similar parameters for ages $> 5\cdot 10^7$ yr. The difference at earlier ages is due to the fact that we assume a differential rotation on the ZAMS.

Varying $\Delta \Omega/\Omega_0$ in Eq. \eqref{eq:omega_ic} in the range $0.01$ to $0.1$, we verified that the amplitude of differential rotation after $\sim 10^7$ yr does not change significantly. This is consistent with a weak dependence on the details of the rotation profile on the ZAMS at later stages.

Our results could be particularly interesting to explain the differences between fast and slow rotators in the core-envelope coupling time-scale, according to the phenomenological models by, e.g.,  \citet{Allain:1998} and \citet{Bouvier:2008b}. If fast rotators reach the MS with a predominatly quadrupolar field, then their coupling time is about ten times shorter than that of slow rotators, that we may assume to have a predominantly dipolar field. The selection of the preferred mode may be determined by the stellar dynamo during the PMS phase \citep[see, e.g., Sect.~3 of][]{Moss_Saar:2008}.

\section{Discussion}

The method used here shares its mathematical foundations with that applied by \citet{Lanza:2007}, where the solution of the PDE expressing the conservation of the angular momentum was expanded in a generalised Fourier series.
In this work, we further develop this approach to solve two coupled PDEs, with non-homogeneous boundary conditions.
This method offers several advantages over finite difference methods, notably the fact that the accuracy  does not degrade with time and that the calculation of the solution at every previous step is not needed if only one particular instant is required.

According to \citet{Spruit:1999},  the strongest instability in a magnetized stellar radiative zone should be that described by \citet{Pitts_Tayler:1985}.
Spruit's analysis was numerically tested by \citet{Zahn_ea:2007}, using a 3D  MHD spectral code.
They confirmed that the strongest non-axisymmetric unstable mode has an azimuthal wavenumber  $m=1$ as for Pitts-Tayler instability, but there was no hint of a regeneration of the poloidal magnetic field owing to a dynamo action associated with the instability, as conjectured by \citet{Spruit:2002}.
Non-axisymmetric motions associated with the instability seem to behave as Alfv\'en waves rather than  turbulence. Their contribution to the transport of angular momentum appears to be negligible,  contrary to the conjectures of \citet{Spruit:1999}, \citet{Denissenkov_Pinsonneault:2007}, and \citet{Denissenkov_Pinsonneault:2008}, who envisaged an increase of the effective turbulent viscosity and diffusivity as a consequence of the motions associated with the instability. Therefore, our assumption of a purely axisymmetric model to study angular momentum transport within a radiative core gains support.

The choice of the initial conditions used in our calculation was mainly motivated by simplicity reasons because no observational constraints on the radial rotational profile or inner magnetic field configuration of stars on the ZAMS is presently available.
Nevertheless, except for  the geometry of the poloidal field, our choice of the initial conditions does not significantly affect  the solution and it is, therefore, not critical for the angular momentum transport problem we have addressed.  

Neglecting the poloidal magnetic field diffusion is acceptable on the basis of the quite long  time-scale of this process.
This could be questionable if some source of enhanced magnetic diffusivity is present in the core, perhaps in association with the still elusive viscosity enhancement, as suggested by \citetalias{Rudiger_Kitchatinov:1996}.
The effect of the diffusion of the poloidal field is merely a decay of its amplitude. This has little influence on the solution and was not taken into account to avoid unnecessary complications.
On the other hand, a possible spatial reshaping of the poloidal field lines, could have a greater impact on the phase-mixing process and the late stages. However, the inclusion of such a field evolution would require a knowledge of the initial magnetic field configuration, which is not available.

Finally we note that a poloidal field in the core may imprint the differential rotation at the outer boundary of the core into the radiative interior of the Sun, contributing to a significant downward spreading of the tachocline which is contrary to helioseismic results \citep[cf. ][]{BrunZahn:2006}. Neglecting the diffusion of the poloidal field and assuming a latitude-independent radial gradient of the angular velocity at the outer boundary, as in our calculations, prevent such an effect. This is equivalent to assume that the  tachocline is not dynamically connected with the core poloidal field, which could be justified if the tachocline is confined into a thin layer above $r=r_{2}$  by other effects, e.g., a highly anisotropic turbulent diffusivity, as proposed by \citet{SpiegelZahn:1992}.  

\section{Conclusions}

We introduce an exact analytic solution of the coupled equations of angular momentum  and azimuthal magnetic field evolution in a radiative core under the main assumptions of axisymmetry and negligible meridional circulation. From such a solution we derive a numerical spectral method that allows us to  address the problem of angular momentum transport in stellar radiative regions. Our illustrative solution of the angular momentum evolution in the radiative core of the Sun is compared to previous numerical models to discuss the advantages of the present approach.
It is particularly intesting because it allows us to  define an analytic upper bound for the numerical errors and it can easily reach hydromagnetic regimes characterized by Reynolds numbers of the order of $10^{7} - 10^{8}$, that are not accessible with other numerical techniques.  Our approach is particularly interesting because it provides a rigorous treatment of the kinetic and magnetic energy distribution among different lenghtscales during the phase mixing process. Previous numerical methods were not capable of a rigorous evaluation of the contribution of the subgrid spatial scales, while our method does it by applying the mathematical closure expressions for generalized Fourier series and the conservation of energy.

With our approach, we studied the evolution of the rotational decoupling of the core, defined through a measure of its differential rotation.
Our results confirm that the uniform rotation of the core of the present Sun, as deduced from helioseismic inversions, can be reproduced only if an enhanced viscosity, $\sim 10^4$ times greater than the molecular value, is assumed. This is an early finding of previous works, but the identification of the underlying physical process(es) is still debated. Asteroseismic measures are needed to infer how general the rigid rotation of the core of the Sun is.

Another interesting result is that a quadrupolar poloidal field leads to a coupling time-scale about one order of magnitude shorter than a dipolar field. We conjecture that such a difference in the geometry of the poloidal field may be a consequence of the different dynamo regimes operating in fast and slowly rotating stars during the PMS stage, respectively. If verified by future studies, this may explain the different coupling time-scales required by the phenomenological models of  rotation braking of fast and slow rotators, respectively. We intend to address further such an interesting topic in a forthcoming work.
Moreover, our analytic treatment of the coupled equations of the angular momentum and magnetic field evolution in an axisymmetric MHD system can be applied to other astrophysically relevant problems, e.g., to study the torsional oscillations of a magnetized shell. This problem has interesting applications to, e.g., the oscillations in roAp stars, as discussed by, e.g., \citet{RinconRieutord03} and \citet{Reeseetal04}. They have considered the case of a magnetized shell of uniform and constant density bathed by a dipolar magnetic field in the absence of rotation, studying both the poloidal and the toroidal axisymmetric modes as well as non-axisymmetric modes by means of linearized equations, i.e., considering only small perturbations. Our approach allows us to study only torsional Alfven waves, i.e., the toroidal modes in their classification, but without the simplifying assumption of small amplitude, and considering also the stratification of the medium and stellar rotation, when the axis of the dipole field is aligned with the rotation axis.
Applications of our approach to strongly magnetized, degenerate stars, such as magnetic white dwarfs or neutron stars \citep[cf., e.g., ][]{OkitaKojima05,Glampedakisetal06}, may also be of interest.
Other possible applications  include an implementation for cylindrically symmetric systems, e.g., an accretion disc  around a protostar threaded by a large-scale dipolar field, to study the torsional Alfven waves and their phase mixing within the ionized region of the disc. However, in this case our model represents an even stronger idealization than in the cases mentioned above  because the geometry of the poloidal field and the large-scale poloidal flow shown by detailed numerical simulations are far from being simply dipolar \citep[cf., e.g., ][]{vonRekowskiBrandenburg04}.   

\section*{Acknowledgements}
AFL wishes to thank Professor J.-P. Zahn for interesting discussion.
The authors thank an anonymous referee for useful comments.
Research on the rotation of the Sun and late-type stars at INAF-Catania Astrophysical Observatory and the Department of Physics
and Astronomy of Catania University is funded by MIUR ({\it Ministero dell'Istruzione, dell'Universit\`a e della Ricerca}), whose financial support is gratefully
acknowledged.
This research has made use of the ADS-CDS databases, operated at the CDS, Strasbourg, France.


\appendix

\section{The wind braking boundary condition}
\label{app:wind}

The amount of shear at the lower boundary of the CZ, assigned through the function $\mathscr{W}(t)$ in Eq. (\ref{eq:om_bc}), can be determined using the integral form of the angular momentum equation applied to the volume of the core $\mathscr{C}$:
\begin{equation*}
\frac{d}{dt} \int_\mathscr{C} \rho r^2 \sin^2 \vartheta \Omega d\tau = \int_\mathscr{C} \nabla \cdot ( \rho r^2 \sin^2 \vartheta \nu \nabla \Omega ) d\tau = \oint_{\mathscr{S}}  \left[ \rho \nu r^2 \sin^2 \vartheta \frac{\partial \Omega}{\partial r} \right]   d \mathscr{S},
\end{equation*}
where $\mathscr{S}$ is the boundary of the core, i.e., the sphere of radius $r_{2}$, and we have applied Gauss theorem, neglecting the azimuthal magnetic stress term $B_{r}B_{\varphi}$ because $B_\varphi =0$ at $r=r_2$ (cf. Eq. \eqref{eq:fl_bc}).

The l.h.s. is the total angular momentum loss suffered by the core. Assuming that $\frac{\partial \Omega}{\partial r}$ on $\mathscr{S}$ does not depend on $\vartheta$ and performing the surface integration, we obtain:
\begin{equation*}
\frac{8 \pi}{3} \rho(r_2) \nu(r_2) r_2^4 \left. \frac{\partial \Omega}{\partial r} \right|_{r_2} = \left( \frac{d J}{d t} \right)_c.
\end{equation*}
\citet{Kawaler:1988} proposed an expression for the angular momentum loss by a magnetised wind, slightly modified by \citet{Chaboyer:1995a, Chaboyer:1995b} to account for the observed \emph{saturation} at high rotation rates:
\begin{equation}
\label{eq:wb}
\left( \frac{d J}{d t} \right)_w  = - K_w \left( \frac{M_*}{M_\odot} \right)^{1/2}\left( \frac{R_*}{R_\odot} \right)^{-1/2} \left\{
\begin{array}{cc}
\Omega_*^3 & \Omega_* \leq \Omega_c \\
\Omega_c^2 \Omega_* & \Omega_* > \Omega_c
\end{array} \right. ,
\end{equation}
where $K_{w} \simeq 10^{47}$ g cm$^2$ s is a constant calibrated in such a way to obtain the present surface angular velocity of the Sun.
The adopted braking law includes a basic dependence on the surface rotation rate $\Omega_*$ by simple power laws, along with a dependence on stellar parameters $R_*$, $M_*$.

To assign the boundary condition in the second of Eqs. (\ref{eq:om_bc}), we equate the flux of angular momentum outside of the core to the angular momentum  lost in the wind as given by Eq. (\ref{eq:wb}):
\begin{equation*}
\left( \frac{d J}{d t} \right)_c \simeq \left( \frac{d J}{d t} \right)_w,
\end{equation*}
In conclusion, we have in the solar case ($M_*=M_\odot$, $R_*=R_\odot$):
\begin{equation*}
\left. \frac{\partial \Omega}{\partial r} \right|_{r_2} = \mathscr{W}(t) = -\frac{3}{8 \pi} \frac{K_w}{\rho(r_2) \nu(r_2) r_2^4} f[\Omega_*(t)],
\end{equation*}
where $f(\Omega_*)$ is the law in the rightmost factor in the r.h.s. of Eq. \eqref{eq:wb}.

The dependence on $\Omega_*(t)$ introduces a non-linearity in the problem. We use the equatorial value of the angular velocity  $\Omega(r_2,\frac{\pi}{2},t)$ to represent $\Omega_*(t)$ for the computation of the boundary condition. Note that we are implicitly assuming that all the  angular momentum lost in the wind is extracted from the boundary of the core. This is equivalent to assume that the CZ and the tachocline can extract angular momentum from the core on a time-scale much shorter than the timescale for the angular momentum redistribution within the core itself. Given the high value of the turbulent viscosity in the CZ this appears to be a plausible hypothesis.

\section{Self-adjoint operators and eigenfunction expansion}
\label{app:selfadj}

Let us consider a linear differential operator of the form:
\begin{equation}
\label{eq:saop}
{\cal L}[\phi] = \nabla \cdot (p \nabla \phi) + q \ \phi,
\end{equation}
with $p$, $q$ given functions of $r$ and $\vartheta$.  

The definition of \emph{self-adjointness} commonly applied to matrices can be generalised for this kind of operators. We introduce the Green formula for the operator $\cal L$, i.e.:
\begin{equation}
\label{eq:green}
\int_{\cal C} \left( u \mathcal{L}[v] - v \mathcal{L}[u] \right) d^3 {\bf r} = \oint_{\cal S} p \left( u \nabla v- v \nabla u \right) \cdot \hat{\bf n} d {\cal S},
\end{equation}
where  $u$ and $v$ are arbitrary differentiable functions and the surface integration extends over the boundary $\cal S$ of the  integration domain $\cal C$, with $\hat{\bf n}$ being the unit vector in the direction of the ourward normal to $\cal S$. It is easily seen that the operator
$\cal L$, as defined by  Eq. \eqref{eq:saop}, satisfies Eq. \eqref{eq:green}.

Let us consider  two arbitrary differentiable functions $\phi$, $\psi$ satisfying linear boundary conditions on $\cal S$ of the kind:
\begin{equation}
\begin{array}{cc}
\alpha \phi + \beta \nabla \phi \cdot {\bf n} &= 0,  \\
\alpha \psi + \beta \nabla \psi \cdot {\bf n} &= 0,
\end{array}
\label{eq:slbc}
\end{equation}
where $\alpha$ and $\beta$ are arbitrary real numbers. Then,  
making use of Eq.~\eqref{eq:green} with $u=\phi$ and $v=\psi$, we find that:
\begin{equation*}
\int_{\cal C} \left( \phi \mathcal{L}[\psi] - \psi \mathcal{L}[\phi] \right) d^3 {\bf r} = 0,
\end{equation*}
which is the defining property of a self-adjoint differential operator.

The eigenvalue problem for a self-adjoint linear operator is specified by the equation:  
\begin{equation*}
{\cal L}[\phi_n] + \kappa_n  w \phi_n =0
\end{equation*}
together with the boundary conditions in Eq. \eqref{eq:slbc}. It is also called a St\"urm-Liouville boundary value problem.
For this kind of problem, an  infinite set of eigenfunctions can be proved to exist, forming a complete and orthogonal set.

The self-adjoint operators appearing in Eqs. \eqref{eq:saom} and \eqref{eq:safl} can be put in the form of Eq. \eqref{eq:saop} with the positions:
\begin{align*}
p &\equiv \rho r^2 \sin^2 \vartheta \nu,  \\
q &\equiv 0,
\end{align*}
for the $Z$ problem, and
\begin{align*}
p &\equiv\eta,  \\
q &\equiv \frac{\eta'}{r} - \frac{\eta}{r^2 \sin \vartheta},
\end{align*}
for the $\Xi$ problem.

We shall now apply the Green formula to derive  Eqs. \eqref{eq:time}. With $u=\Omega$ and $v=Z$, Eq. \eqref{eq:green} becomes:
\begin{equation*}
\int_\mathscr{C} \left[ \Omega \nabla \cdot (\rho r^2 \sin^2 \vartheta \nu \nabla Z_{nk}) - Z_{nk} \nabla \cdot (\rho r^2 \sin^2 \vartheta \nu \nabla \Omega) \right] d^3 {\bf r} = \oint_\mathscr{S} \rho r^2 \nu \sin^2 \vartheta \left( \Omega \frac{\partial Z_{nk}}{\partial r} - Z_{nk} \frac{\partial \Omega}{\partial r} \right) d \mathscr{S},
\end{equation*}
where the integration is extended over the computational domain $\mathscr{C} \equiv [r_1,r_2]\times[-1,+1] \times [0, 2 \pi]$, bounded by the surface $\mathscr{S} \equiv S_1 \cup S_2$, where $S_1$ and $S_2$ are the two spherical surfaces of centre $O$ and radius $r_1$, $r_2$, respectively.
This formula allows us to  take into account the boundary conditions. Using Eqs. \eqref{eq:zbc} and \eqref{eq:om_bc}, the r.h.s. becomes:
\begin{equation*}
\oint_\mathscr{S} \rho r^2 \nu \sin^2 \vartheta \left( \Omega \frac{\partial Z_{nk}}{\partial r} - Z_{nk} \frac{\partial \Omega}{\partial r} \right) d\Sigma = - 2 \pi \rho(r_2) \nu(r_2) r_2^4 \zeta_{nk}(r_2) \mathscr{W}(t) \int_{-1}^{+1} \sin^2 \vartheta P^{(1,1)}_n d\cos \vartheta \equiv - \mathscr{W}_{nk}(t).
\end{equation*}
The left hand side of the Green formula can be further manipulated by using Eqs. \eqref{eq:saom} and \eqref{eq:om_sc}, i.e. with the substitutions:
\begin{eqnarray*}
\nabla \cdot (\rho r^2 \sin^2 \vartheta \nu \nabla Z_{nk}) &=& - \rho r^2 \sin^2 \vartheta \lambda_{nk} Z_{nk}, \\
\nabla \cdot (\rho r^2 \sin^2 \vartheta \nu \nabla \Omega) &=& \mathscr{R}_\nu \left[ \rho r^2 \sin^2 \vartheta \frac{\partial \Omega}{\partial t}  - {\bf B}_p \cdot \nabla (r \sin \vartheta B_\varphi) \right].
\end{eqnarray*}
Recalling that $Z_{nk}(r,\vartheta) = \zeta_{nk}(r) P^{(1,1)}_n(\vartheta) $, we obtain:
\begin{eqnarray*}
\int_\mathscr{C} \Omega \nabla \cdot (\rho r^2 \sin^2 \vartheta \nu \nabla Z_{nk}) d^3r &=& -\lambda_{nk} \int_{-1}^{+1} \int_{r_1}^{r_2} \rho r^2 \sin^2 \vartheta \zeta_{nk} P^{(1,1)}_n \Omega r^2 dr\ d\cos \vartheta \equiv - \lambda_{nk} \omega_{nk} \\
\int_\mathscr{C} Z_{nk} \nabla \cdot (\rho r^2 \sin^2 \vartheta \nu \nabla \Omega) d^3r &=& \mathscr{R}_\nu \left[ \int_{-1}^{+1} \int_{r_1}^{r_2} \rho r^2 \sin^2 \vartheta \zeta_{nk} P^{(1,1)}_n \frac{\partial \Omega}{\partial t} r^2 dr\ d\cos \vartheta \right. \\
&-&\left. \int_{-1}^{+1} \int_{r_1}^{r_2} \rho r^2 \sin^2 \vartheta \zeta_{nk} P^{(1,1)}_n {\bf B}_p \cdot \nabla (r \sin \vartheta B_\varphi) r^2 dr\ d\cos \vartheta \right] \\
&\equiv& \mathscr{R}_\nu \left( \frac{d \omega_{nk}}{dt} - \sum_{mh} S_{nkmh} \beta_{mh} \right).
\end{eqnarray*}
The last two equalities were established using the definition of the generalised Fourier coefficients in Eqs. \eqref{eq:omcoeff} and \eqref{eq:flcoeff} and their time derivatives. Moreover, the linear dependence of the source term on $B_\varphi$ allows us to express the spatial dependence of its expansion through  suitable coefficients $S_{nkmh}$, entering as factors of $\beta_{mh}(t)$ and whose explicit expression is given below.
Thus, by applying the Green formula  to Eq. \eqref{eq:saom},
provided that the eigenfunctions $Z_{nk}$ are conveniently normalised (i.e. $\left<Z_{nk},Z_{nk} \right> \equiv 1$), we obtain the first of Eqs. \eqref{eq:time}:
\begin{equation*}
- \lambda_{nk} \omega_{nk} -  \mathscr{R}_\nu \left( \frac{d \omega_{nk}}{dt} - \sum_{mh} S_{nkmh} \beta_{mh} \right) = - \mathscr{W}_{nk}.
\end{equation*}
Applying the same line of reasoning, the derivation of the second of Eqs. \eqref{eq:time} is quite straightforward. The boundary terms vanish completely in view of Eqs. \eqref{eq:fl_bc} and \eqref{eq:zbc} and we are left with:
\begin{equation*}
\int_\mathscr{C} \left\{ B_\varphi \mathcal{D} [ \Xi_{nk} ] - \Xi_{nk} \mathcal{D} [ B_\varphi ]  \right\} d^3 {\bf r} = 0,
\end{equation*}
with the shorthand notation: $\mathcal{D}[\Xi_{nk}] \equiv \nabla \cdot (\eta \nabla \Xi_{nk}) - \left[ \dfrac{\eta'}{r} - \dfrac{1}{r^2 \sin^2 \vartheta} \right] \Xi_{nk}$.

Substitution of the eigenvalue problem, Eq. \eqref{eq:safl}, and of the PDE, Eq. \eqref{eq:fl_sc},
i.e.,
\begin{eqnarray*}
{\cal D}[ \Xi_{nk}] &=& -\mu_{nk} \Xi_{nk}, \\
{\cal D}[B_\varphi] &=& \mathscr{R}_\eta \left( \frac{\partial B_\varphi}{\partial t} - r \sin \vartheta \nabla \Omega \right),
\end{eqnarray*}
leads to:
\begin{eqnarray*}
&-& \mu_{nk} \int_{-1}^{+1} \int_{r_1}^{r_2} \xi_{nk} P^{1}_n B_\varphi r^2 dr\ d\cos \vartheta \\
&-& \mathscr{R}_\eta \left( \int_{-1}^{+1} \int_{r_1}^{r_2} \xi_{nk} P^{1}_n \frac{\partial B_\varphi}{\partial t} r^2 dr\ d\cos \vartheta - \int_{-1}^{+1} \int_{r_1}^{r_2} \xi_{nk} P^{1}_n r \sin \vartheta {\bf B}_p \cdot \nabla \Omega r^2 dr\ d\cos \vartheta \right) \\
&=& -\mu_{nk} \beta_{nk} - \mathscr{R}_\eta \left( \frac{d \beta_{nk}}{dt} - \sum_{mh} T_{nkmh}\omega_{mh} \right)=0,
\end{eqnarray*}
which is indeed equivalent to the second of Eqs. \eqref{eq:time}.

For reference, we provide here the complete expressions of $S_{nkmh}$ and $T_{nkmh}$, omitting the huge amount of algebra necessary to derive them:
\begin{eqnarray*}
S_{nkmh} &=& \int_{-1}^{+1} \chi \frac{d }{d x}\left[ \sqrt{1-x^2} P^1_m\right] P^{(1,1)}_n dx \cdot \int_{r_1}^{r_2} r \frac{d \tilde{\varphi}}{dr}  \xi_{mh}  \zeta_{nk} dr  \\\nonumber
&-&  \int_{-1}^{+1} \sqrt{1-x^2} \frac{d \chi}{d x} P^{(1,1)}_n P^1_m dx \cdot \int_{r_1}^{r_2} \tilde{\varphi} \frac{d}{dr}\left[ r \xi_{mh} \right] \zeta_{nk} dr  
\\
T_{nkmh} &=&  \int_{-1}^{+1} \chi \sqrt{1-x^2} \frac{d  P^{(1,1)}_m}{d x}  P^1_n dx \cdot \int_{r_1}^{r_2} r \frac{d \tilde{\varphi}}{dr}  \xi_{nk}  \zeta_{mh} dr    \\\nonumber
&-&  \int_{-1}^{+1} \sqrt{1-x^2} \frac{d \chi}{d x} P^{(1,1)}_m P^1_n dx \cdot \int_{r_1}^{r_2} r \tilde{\varphi} \frac{d \zeta_{mh}}{dr} \xi_{nk} dr  
\end{eqnarray*}
where the functions $\tilde{\varphi}$ and $\chi$ appear in the expression of the poloidal field flux function, i.e., $\Psi = \tilde{\varphi}(r) \chi (x)$, with $x = \cos \vartheta$.

From these expressions, a remarkable property of the matrix $\mathbb{A}$ in Eq. \eqref{eq:odecomp} can be proved. Using the boundary conditions  for the eigenfunctions $\xi_{nk}$ in Eqs. \eqref{eq:zbc} and integrating by parts, we find:
\begin{equation}
\label{eq:A_symm}
T_{nkmh}=-S_{mhnk}.
\end{equation}
With the notation ${\bf L} = \rm{diag}(-\lambda_{j}/\mathscr{R}_\nu)$,  ${\bf M} = \rm{diag}(-\mu_{j}/\mathscr{R}_\eta)$, $\mathbb{S}=(S_{jl})$, where $j$, $l$ are indexes that number the orderer couples $(n,k)$, $(m,h)$, respectively, e.g., $j=Kn+k$, $l=Km + h$, we can write:
\begin{equation*}
\mathbb{A} = \left(
\begin{array}{c|c}
{\bf L} & \mathbb{S} \\
\hline
-\mathbb{S}^T & {\bf M}
\end{array}
\right) =
\left(
\begin{array}{c|c}
{\bf L} & 0 \\
\hline
0 & {\bf M}
\end{array}
\right) +
\left(
\begin{array}{c|c}
0 & \mathbb{S} \\
\hline
-\mathbb{S}^T & 0
\end{array}
\right).
\end{equation*}
Thus we see that $\mathbb{A}$ is the sum of a diagonal matrix plus an antisymmetric matrix. Since both diagonal and antisymmetric matrices commute with each other and with their respective transposes, i.e., they are \emph{normal matrices}, also their sum $\mathbb{A}$ is a normal matrix.
This is a very useful property, because for normal matrices a generalisation of the spectral theorem holds, which is then applicable to Eq. \eqref{eq:odecomp}, as discussed in Sect.~\ref{complete_eqs}.

\label{lastpage}

\end{document}